\documentclass[a4paper,fleqn]{cas-sc}

\usepackage[numbers,sort&compress]{natbib}

\usepackage{tikz}
\usepackage{subcaption}
\usepackage{float}
\usepackage{grffile}
\usepackage{booktabs} \usepackage{siunitx}
\usepackage{gensymb}
\usepackage{tabularx}

\begin{document}

\shorttitle{Procedural Volumetric Modeling of Plant Branching Structures for Finite Element Analysis}
\shortauthors{A. Moola et~al.}

\title[mode=title]{Procedural Volumetric Modeling of Plant Branching Structures for Finite Element Analysis}

\author[1]{Ajith Moola}
\author[1]{Prashant Kumar Gupta}
\author[1]{Baskar Ganapathysubramanian}
\author[1]{Aishwarya Pawar}

\affiliation[1]{organization={Department of Mechanical Engineering, Iowa State University},
country={United States}}

\begin{abstract}
Precision agriculture, smart breeding, and agricultural robotics require accurate and automated plant modeling. These models provide high-fidelity three-dimensional (3D) representations of plant architecture. They provide the geometric foundation for simulations of water and nutrient transport, light interception, structural loading, and crop lodging. Unlike static plant modeling pipelines, procedural modeling frameworks not only generate accurate 3D plant geometries but also support the generative modeling of crop diversity and the dynamic modeling of plant growth. While terrestrial laser scanning, LiDAR, photogrammetry, and neural reconstruction-based approaches have made 3D plant reconstruction possible, the resulting data are typically in the form of point clouds, which cannot be directly utilized for high-fidelity simulations. We present an automated volumetric procedural modeling framework for plant branching structures that generates analysis-suitable hexahedral meshes from input skeletons or 3D point clouds. The input skeleton is first converted into a rooted graph representation that captures the plant branching topology. Each graph edge is then represented by a smooth centerline B-spline curve, around which a cylindrical tensor-product B-spline volume is constructed. At each junction, incident B-spline volume control lattices are joined using blending operations. The resulting volumetric parameterization is evaluated to generate a smooth and conforming hexahedral mesh of the whole plant. We demonstrate the framework on three diverse plant datasets, namely mung bean, tomato, and walnut trees, generating meshes with both uniform and spatially varying branch radii. Mesh quality is evaluated using the element-wise scaled Jacobian metric, and solver readiness is demonstrated by solving a steady-state diffusion problem on the generated meshes. The framework also supports dynamic mesh generation suitable for modeling plant growth by locally updating newly added branches without reconstructing the full plant geometry. This procedural framework thus bridges the gap between skeleton-based plant geometry and analysis-suitable volumetric modeling for plant biomechanics, transport simulations, and digital twin applications.
\end{abstract}


\maketitle
\newcounter{algctr}
\newenvironment{algbox}[1]{%
  \refstepcounter{algctr}%
  \par\medskip\noindent
  \begin{minipage}{\linewidth}%
  \hrule height 0.7pt\smallskip
  \textbf{Algorithm \thealgctr.} #1\par
  \hrule height 0.4pt\smallskip
  \begin{enumerate}\setlength{\itemsep}{2pt}\setlength{\parsep}{0pt}\setlength{\topsep}{0pt}%
}{%
  \end{enumerate}
  \smallskip\hrule height 0.7pt
  \end{minipage}\par\medskip%
}

\setlength{\textfloatsep}{6pt plus 1pt minus 1pt}
\setlength{\intextsep}{6pt plus 1pt minus 1pt}

\section{Introduction}
Accurate geometric modeling of plant architecture is a fundamental requirement in computational agricultural applications, such as digital twins \cite{ganapathysubramanian2025digital, mitsanis20243d, escriba2024digital,  subeesh2025agricultural}, functional-structural plant models \cite{zhang2026overview, chang2023geometric, sievanen2014functional, soualiou2021functional}, precision agriculture \cite{stansluos2024planting, deng2012insights}, resource optimization \cite{yan2004dynamic, rosell2012review}, genotype-to-phenotype modeling \cite{xu20263d, struik2016bridging, ziamtsov2020plant}, bio-design \cite{bucksch2017morphological, pradal2009plantgl, fourcaud2008plant}, and crop breeding \cite{hadadi2025procedural, yang2026three, zhang2023crop, de2009coupling}. High-fidelity geometric models generated from advanced sensor datasets provide quantitative analysis of plant architecture traits under different genetic, environmental, and management conditions, improving agricultural productivity and resilience \cite{omia2026advancements, rosell2012review, guo2011plant, soualiou2021functional}. By integrating real-time sensor data and environmental parameters, plant models can provide a realistic virtual representation of plants, enabling non-invasive, dynamic, and bi-directional decision-making. In precision agriculture, they allow for accurate morphological evaluations, canopy architecture design, efficient plant breeding, and resource usage \cite{azrai2025canopy, yang2026three}. In simulation-driven crop design, high-fidelity crop models can be used in inverse design applications to generate idealized ideotypes for high productivity and improved resilience \cite{christensen2018use, saleem2025algorithmic, zavorskas2024incremental, jubery2019silico, carbajal2024role}. The goal is not only to generate plant models for visualization but also to use them directly for physics-based simulations, such as nutrient and water uptake, structural mechanics \cite{wang2023finite, sayad2023semi, bidhendi2018finite, jackson2019fea}, light-use efficiency, and ecophysiological and ecosystem modeling \cite{yu2023crop, paulus2019measuring, bomer20243d, wang2024functional, deb2025biophysical, prusinkiewicz2000simulation, zheng2025mechanistic}. 

The advent of advanced sensor technologies for three-dimensional (3D) plant reconstruction, such as light detection and ranging (LiDAR), terrestrial laser scanning (TLS), multiview stereo (MVS), structured light scanning, neural radiance fields (NeRF), and Gaussian splatting, has enabled non-invasive, high-throughput plant phenotyping and monitoring \cite{ghose2026advancements, song2025comprehensive, dietrich2025advancing, liu2025survey, harandi2023make, bomer20243d, xiao2022advanced}. Whole plant architecture is typically represented as depth maps, point clouds, or surface meshes. These reconstructions provide high-resolution quantitative and qualitative visual representations of plant architecture, but they are not directly suitable for physics-based simulations \cite{sankaramaddi20262d, joshi2026neural, hadadi2026floraforge, hadadi2025procedural, xie2023generating, rivera2023lidar}. This identifies a critical research gap, where we have access to large datasets of high-fidelity 3D plant reconstructions acquired from sensors. However, converting them into analysis-ready volumetric meshes remains computationally intensive, requires manual effort, and is not scalable for complex geometries and challenging applications.

Procedural modeling approaches encode plant geometry in a rule-based manner, mathematically embedding the latent plant branching architecture and topology. Grammar-based approaches, such as L-systems, encode branching topology and plant growth using string-based rules \cite{boudon2012py, prusinkiewicz2018modeling,herulambang20163d, leitner2010dynamic, lee2023latent, bernard2021techniques, allen2004using, ong2014approach}. Functional-structural plant models (FSPMs) utilize algorithmic rules to construct plant geometry and employ physiological processes to guide the development of plant morphology \cite{wang2024functional, dejong2011using, feng2014comparing, zhang2026overview, vos2010functional, louarn2020two, sievanen2014functional, du2025hug}. Non-uniform rational B-splines (NURBS) have been used in procedural modeling frameworks to accurately generate high-fidelity plant representations \cite{hadadi2026floraforge, hadadi2025procedural, wu2021parametric, ando2021robust, murata2024three}. These approaches allow for biologically realistic plant modeling, but they are limited to implicit or surface representations, which require considerable computational effort to generate analysis-suitable volumetric meshes.

Generating analysis-suitable volumetric meshes for branching structures is an especially challenging and non-trivial problem, particularly near junctions, where multiple branch segments connect at different angles. A robust volumetric modeling framework should capture the correct plant branching topology for all possible configurations while generating a regular and conforming mesh that is free of intersections, gaps, and distortions \cite{cheikh2025tube2fem, pietroni2022hex}. Conventional volumetric meshing and NURBS-based modeling workflows for generating branching geometries can create highly accurate and high-quality models \cite{li2023intracellular, zhang2007patient, li2023isogeometric, li2022modeling, li2022modelingJ}. However, these approaches are static; modifying the branching topology, such as adding or deleting branches or changing the branch radius, requires considerable effort to remesh or even rebuild the entire geometry from scratch. This highlights a key limitation: there are currently no procedural volumetric meshing frameworks for automatically generating branching geometries while remaining compatible with changing plant topology.

We present an automated volumetric procedural modeling framework for generating analysis-suitable hexahedral (hex) meshes of plant branching structures from 3D point clouds and their corresponding plant skeletons \cite{godin1999mtg, pfeifer2004trees, cornea2007curve, cao2010skeletons, huang2013l1, tagliasacchi2016skeletons, middleton2022representing}. The input skeleton is first converted into a rooted directional graph representation that captures the plant topology. B-spline curves are then fitted to each branch, rendering a smooth centerline geometry. Rotation-minimizing frames along the centerline curves are used to construct cylindrical B-spline volumes for individual branches. At bifurcation and trifurcation junctions, incident branch B-spline volumes are joined via control-point blending operations, enforcing seamless transitions. The resulting volumetric parameterization of the complete plant is then evaluated to generate a conforming and smooth hex mesh suitable for finite element analysis (FEA). 

The proposed framework utilizes a compact set of rule-based generation steps instead of manual mesh construction. It generates plant geometry through a sequence of steps, progressing from graph and curve representations to volumetric meshes. The framework not only enables automatic volumetric mesh generation from plant skeleton data but also allows local updates without reconstructing the whole plant geometry as the plant topology changes. The main contributions of our work are:
\begin{enumerate}
    \item A procedural volumetric modeling framework that automatically converts 3D plant skeleton data obtained from sensors into hex meshes suitable for FEA.
    \item A branch-wise B-spline volumetric parameterization strategy that first represents each branch centerline as a B-spline curve. Rotation-minimizing frames are constructed along the centerline curves, and tensor-product cylindrical B-spline volumes are generated for each branch.
    \item A junction-blending procedure is applied to bifurcations and trifurcations, joining incident branch volumes with seamless transitions and producing conforming volumetric meshes for the entire plant geometry.
    \item Demonstration of analysis-suitable volumetric mesh generation for diverse plant species of increasing geometric complexity, which includes mung bean, tomato, and walnut trees. 
    \item Robust generation of plant hex meshes  with spatially varying branch radii from unstructured 3D point clouds.
    \item Demonstration of dynamic mesh generation for plant growth modeling, where newly added branches are locally parameterized and blended without reconstructing the whole plant geometry.
\end{enumerate}

\section{Related work}
In recent years, there has been significant advancement in the field of 3D plant reconstruction technologies, enabling the capture of detailed structural and morphological plant traits for accurate phenotyping. These include active depth-based imaging technologies such as LiDAR \cite{maeda2025expanding},  TLS \cite{medic2023challenges, rodriguez2024cotton, liang2024advanced, bekkers2025improving}, structured-light sensors \cite{nguyen2015structured, wang2020research}, and time-of-flight cameras \cite{song2023dynamic}, which use light or infrared signals to generate depth maps. Passive 3D reconstruction methods, such as stereo vision, MVS \cite{wang2025accurate, daiki2024comparative, rose2015accuracy,martinez2019low}, and  structure-from-motion(SfM) photogrammetry \cite{fernandez2022estimation}, offer ways to reconstruct depth maps without emitting signals. However, these methods generate point clouds that suffer from noise, occlusions, and missing data \cite{omia2026advancements}. More recently, deep learning-based approaches such as NeRF \cite{choi2024nerf, kang2023high} and 3D Gaussian splatting \cite{zhang2025wheat3dgs} have emerged as phenotyping tools that generate high-resolution point clouds \cite{arshad2024evaluating, joshi2026drone, joshi2026neural,stuart2025high, wu20253d}. These approaches generate continuous volumetric scene representations, allowing them to handle thin structures and occlusions more easily \cite{ghose2026advancements, li2025survey}. After 3D point cloud acquisition, the phenotyping process proceeds with image analysis steps, such as point cloud registration \cite{chebrolu2021registration}, surface reconstruction \cite{stausberg20243d, hadadi2025procedural, kimara2025maizefield3d, ando2021robust}, semantic segmentation  \cite{zarei2024plantsegnet}, and skeletonization \cite{wu2019accurate, chaudhury2020skeletonization}, to extract key morphological traits \cite{wang2020research, harandi2023make}, yielding segmented point clouds, voxels, surfaces, or skeletons that describe plant geometry. Current approaches focus primarily on surface reconstruction for plant phenotyping, using triangulations \cite{jay2015field, gelard20163d}, B-splines, or NURBS \cite{santos20143d, zhuo2026mirage} to represent plant architecture at the organ, plant, or canopy scales \cite{ hadadi2026floraforge, kimara2025maizefield3d}. As a result, analysis-suitable volumetric modeling remains largely absent from plant phenotyping.

Procedural plant modeling frameworks utilize algorithmic rules rather than manual geometry generation to model 3D plant architecture. Using a small set of geometric parameters and a rule-based generation approach, geometric models can be generated more efficiently for realistic plant growth and phenotyping applications \cite{stava2014inverse, li2024interactive}. Lindenmayer Systems (L-systems) are rule-based procedural modeling approaches used to generate realistic 3D plant structures through a recursive algorithmic generation of L-strings that encode plant topology, geometry, and growth \cite{ prusinkiewicz1990abop}. Deterministic, stochastic, and parametric L-systems have been shown to model realistic 1D plant branching architectures and growth processes for plant visualization and botany \cite{prusinkiewicz1998modeling,prusinkiewicz2018modeling, cieslak2019gillespie}. Subsequent research extended these approaches to include interactive and sketch-based tree modeling and procedural reconstruction from images or point clouds \cite{guo2020inverse, lee2023latent, magnusson2023towards, bernard2023stochastic}. These approaches showcase the flexibility of the rule-based frameworks for both forward and inverse data-driven reconstruction. FSPM approaches utilize physiological processes along with plant geometry to model plant growth and function \cite{vos2010functional, evers2018computational, sievanen2014functional}. The geometric modeling components, such as plant organs and internodes, are combined with mechanistic modules for photosynthesis, carbon allocation, and resource competition in a unified simulation framework \cite{louarn2020two, wang2024functional, o2021integrating, bekkers2025improving, soualiou2021functional}. Both the L-system and FSPMs provide geometric outputs for visualization, trait analysis, and growth simulation; however, they cannot be directly used for physics-based simulations \cite{vos2010functional, evers2018computational}.

Recently, data-driven approaches have utilized machine learning to encode key morphological parameters, such as plant topology and geometry, into compact representations \cite{debbagh2024generative, zhai2024cropcraft, liu2021treepartnet, wu20253d}. By training on large imaging and point cloud datasets, supervised and deep learning approaches, such as convolutional neural networks (CNNs) and multi-layer perceptrons (MLPs), recover parameters for morphological traits and plant growth, enabling automated trait estimation and complementing rule-based or mechanistic models \cite{liu2021treepartnet}. They also provide low-dimensional embeddings of complex plant topology and geometry for inverse modeling and generative modeling of diverse plant geometries under varying environmental and genetic conditions. Computer-aided design (CAD) modeling allows for a smooth parametric representation of complex geometry. Spline-based modeling allows for precise control, enabling both local and global high-fidelity shape representation \cite{zhuo2026mirage}. B-spline curves or generalized cylinders have been used to represent plant branches, crowns, roots, and plant organs \cite{wen2018leaf, okura20223d, aakerblom2015analysis}. A generalized cylinder-based approach for modeling plant branches with implicit ramiform junctions was introduced in \cite{bloomenthal1985maple}. In \cite{wu2009interactive}, plant branches were represented as 3D B-spline curves coupled with a 1D B-spline radius; however, junction blending was identified as an open problem. Generalized B-spline crown envelopes have been used to model the complex polymorphism of Chinese fir to support natural pruning simulations \cite{cui2023constructing}. B-splines have also been coupled with L-systems for the visual modeling of rice root growth \cite{yang2022visual, yang2020modeling}. Recent methods utilize B-spline centerline approaches along with parallel transport frames for plant geometric modeling. Procedural frameworks for NURBS-based surface fitting to point cloud data have been developed, with clear gaps remaining in analysis-suitable volumetric modeling \cite{hadadi2026floraforge, hadadi2025procedural, moola2024thb, moola2026valvefit, prasad2022nurbs, du2025hug}.

Volumetric hex meshing and spline-based volumetric parameterization (V-rep) methods have been developed for FEA and isogeometric analysis (IGA) for branching structures \cite{hughes2005iga, cottrell2009isogeometric, li2022modeling,
li2021deep, li2023isogeometric, wei2022analysis, joshi2022building, joshi2020development, yu2020anatomically, zheng2022volumetric, sederberg2003tsplines, liu2014vtspline, collin2016asg1, bazilevs2006fsi}. Trivariate tensor-product B-spline volumes were fitted to genus-0 surfaces using harmonic parameterization in the cylindrical coordinate system in \cite{martin2008volumetric}. In \cite{massarwi2016b}, trimmed tri-variate B-spline volumes using geometric primitives were glued along shared boundaries with $C^{0}$-continuity. Patient-specific NURBS volumetric modeling has been carried out for vascular structures using skeleton-based sweeps with template-based junction continuity enforcement \cite{zhang2007patient}. Polycube and block-decomposition  approaches have been used to generate high-quality, analysis-suitable volumetric models for complex geometries \cite{yu2014optimizing, yu2022hexdom, yu2022hexgen, yu2025dl, kowalski2012fun, fang2016all,gregson2011all, hu2016centroidal}. These robust, analysis-suitable modeling approaches, however, are not tailored for plant phenotyping workflows and typically assume static geometric construction, with limited applicability for modeling dynamic plant branching topologies.

\section{Methods}
We present a procedural framework that generates hex meshes for FEA from input plant skeleton point clouds. These point clouds approximate the medial axis of the plant, capturing the branching topology and centerline geometry. We denote the 3D point cloud by $\mathbf{X}=\{\mathbf{x}_l\}_{l=1}^{N}$, with $\mathbf{x}_l\in\mathbb{R}^3$, where $N$ denotes the total number of points. The input point cloud is first converted into a skeleton graph that encodes the branching topology and connectivity. The skeleton graph is then transformed into an analysis-suitable hex mesh through five steps: (1) extraction of a branch graph from the skeleton graph by collapsing maximal chains of degree-two vertices; (2) fitting a B-spline centerline curve to the coordinate sequence along each edge of the branch graph; (3) construction of rotation-minimizing reference frames along each branch centerline curve by parallel transport; (4) construction of a hollow cylindrical tensor-product B-spline volume for each branch; and (5) blending of control points at each junction node. The B-spline-based volumetric parameterization provides a smooth, locally controllable representation of the branch geometry. The compact support of B-spline basis functions confines junction blending to a small number of control point layers in the volume. The output of the framework is a conforming hex mesh obtained by evaluating and assembling the blended B-spline branch volumes. The detailed procedure for each of these steps is described in the remainder of this section.

\subsection{Skeleton and branch graph representation}
\label{subsec:tree-representation}
Let $\mathcal{G}_S = (\mathcal{V}_S, \mathcal{E}_S)$ denote the \emph{skeleton graph}, with nodes $\nu_l \in \mathcal{V}_S$ and edges $e_{m} \in \mathcal{E}_S$. We assume that the input skeleton graph has been cleaned to form a connected graph that captures the correct branching topology; skeleton extraction and cleanup from the raw point cloud are outside the scope of this work. Each node $\nu_l$ is associated with an embedded coordinate $\mathbf{x}_l$ and a non-negative scalar $R_{l} \ge 0$ representing the outer radius of the cross-section at that coordinate. From $\mathcal{G}_S$, we then construct a \emph{branch graph} $\mathcal{G}_B = (\mathcal{V}_B, \mathcal{E}_B)$ that is treated as a rooted tree, with all edges oriented from the root toward the leaves. The node set $\mathcal{V}_B \subseteq \mathcal{V}_S$ is defined by retaining all nodes in $\mathcal{V}_S$ whose degree is not equal to two:
\begin{equation}
\mathcal{V}_B = \{\nu_l \in \mathcal{V}_S \mid \deg_{\mathcal{G}_S}(\nu_l) \neq 2\}.
\end{equation} Here, the degree of the node, $\deg_{\mathcal{G}_S}(\nu_l)$, is the number of edges in $\mathcal{G}_S$ incident to $\nu_l$. Nodes with $\deg_{\mathcal{G}_S}(\nu_l) = 1$ are terminal nodes. Among the terminal nodes, we select the root node, denoted by $\nu_{\mathrm{root}}$, with $\mathcal{V}^{\mathrm{root}}_B=\{\nu_{\mathrm{root}}\}$ associated with either the basal skeleton vertex of the point cloud, such as the vertex corresponding to the lowest $z$-coordinate, or a user-defined vertex. The remaining terminal nodes form the leaf set $\mathcal{V}^{\mathrm{leaf}}_B$. Nodes with $\deg_{\mathcal{G}_S}(\nu_l) \geq 3$ are the junction nodes. Every junction node in the junction set $\mathcal{V}^{\mathrm{junc}}_B$ has one incoming edge and at least two outgoing edges in $\mathcal{G}_B$. The node set $\mathcal{V}_B$ can thus be partitioned into the root, junction, and leaf sets as
\begin{equation}
\mathcal{V}_B
=
\mathcal{V}^{\mathrm{root}}_B
\;\dot{\cup}\;
\mathcal{V}^{\mathrm{junc}}_B
\;\dot{\cup}\;
\mathcal{V}^{\mathrm{leaf}}_B.
\label{eq:node_partition}
\end{equation} Each directed edge from a proximal node $\nu_\beta$ to a distal node $\nu_{\alpha}$ is denoted by $b_{\alpha} \in \mathcal{E}_B$. We identify each edge $b_{\alpha}$ by its distal node $\nu_{\alpha}$. Each edge $b_{\alpha}$ stores an ordered coordinate sequence $\mathbf{Q}^{\alpha} = (\mathbf{q}^{\alpha}_1, \dots, \mathbf{q}^{\alpha}_{M^{\alpha}})$, corresponding to the unique maximal path in 
$\mathcal{G}_S$ from $\nu_\beta$ to $\nu_{\alpha}$, whose interior nodes all have degrees equal to two. The endpoints satisfy $\mathbf{q}^{\alpha}_1=\mathbf{x}_{\beta}$ and $\mathbf{q}^{\alpha}_{M^{\alpha}}=\mathbf{x}_{\alpha}$, where $M^{\alpha}$ denotes the number of skeleton points along each edge $b_{\alpha}$.

\begin{figure}[pos=htbp]
\centering\includegraphics[width=0.97\linewidth]{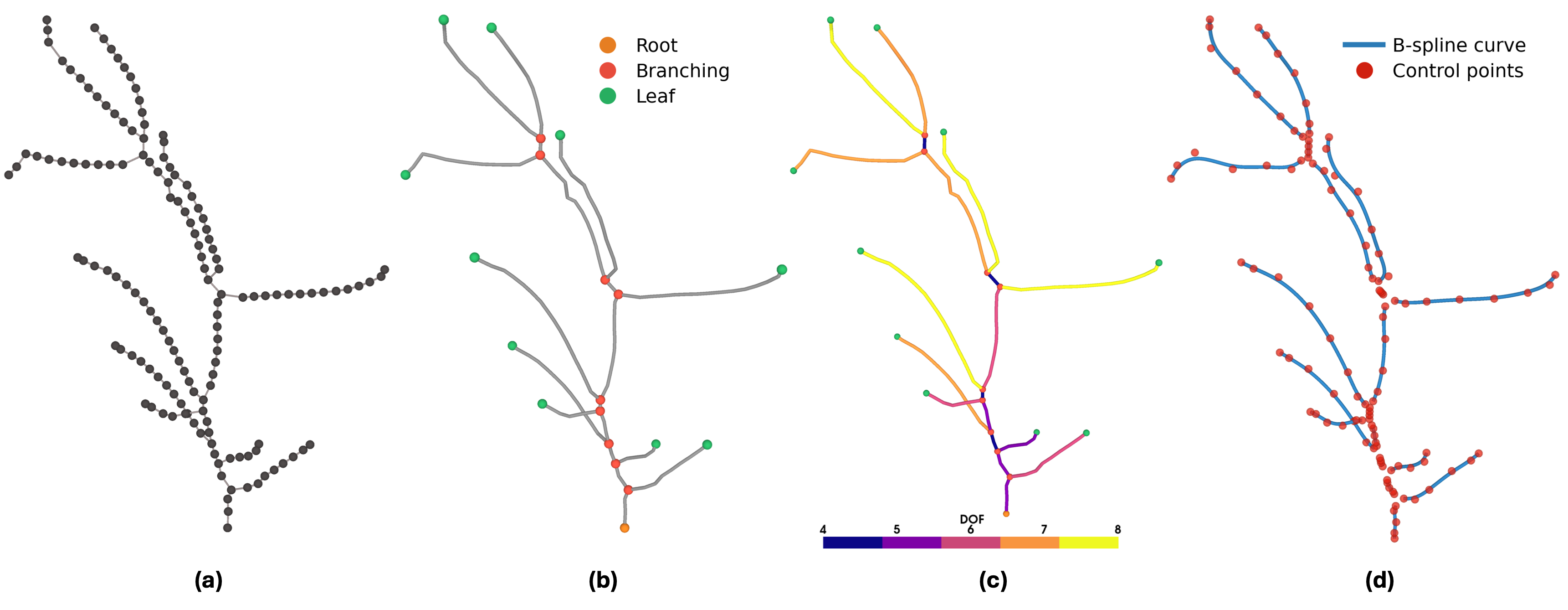}
    \caption{\textbf{Overview of the skeleton-to-centerline curve pipeline.} \textbf{(a)} Skeleton graph of a mung bean plant. \textbf{(b)} Branch decomposition of the skeleton graph into a directed branch graph where each node is classified as a leaf node (green), junction node (red) or root node (orange) and each edge represents an individual branch. \textbf{(c)} Edges in the branch graph colored according to their assigned degrees of freedom (DOF), corresponding to the number of control points used for each B-spline centerline curve. \textbf{(d)} Fitted B-spline centerline curves for each branch along with their corresponding control points.}
\label{fig:skeleton_to_curve}
\end{figure}

\subsection{B-spline centerline curve construction}
\label{subsec:centerline-curve-construction}
We represent the centerline geometry of each edge $b_{\alpha} \in \mathcal{E}_B$ by a smooth centerline B-spline curve, $\mathbf{C}^{\alpha}:[0,1]\rightarrow\mathbb{R}^3$, defined as: 
\begin{equation}
    \mathbf{C}^\alpha(u) \;=\; \sum_{k=1}^{n^\alpha} N_{k,p}(u)\, \mathbf{P}^\alpha_k, \qquad u\in[0,1],
    \label{eq:centerline}
\end{equation} where $n^\alpha$ denotes the number of control points, $N_{k,p}(u)$ denotes the $k^{th}$ univariate basis function of degree $p$ defined on the clamped knot vector $\mathcal{U}^{\alpha} := \{u_1, u_2, \dots, u_{n^{\alpha} + p + 1}\}$, and $\mathbf{P}^\alpha_k \in \mathbb{R}^3$ is the corresponding B-spline control point \cite{piegl_nurbs_1995}. We present two strategies to construct the centerline curves for the entire branch graph. The global construction strategy (Section~\ref{subsubsec:global-construction}) extracts the complete branch graph and fits an individual B-spline curve to each edge $b_{\alpha}$. In the hierarchical construction strategy (Section~\ref{subsubsec:hierarchical-construction}), the branch graph is initialized with a primary path (main stem) in the skeleton graph from the root node to a terminal leaf node. A B-spline curve is first fitted to this primary path to represent the centerline geometry. The complete branch graph is then grown recursively by attaching child branches one at a time. At each new junction, a knot insertion operation is performed on the B-spline curve to split the parent branch, create space for attaching the child branch, and enforce junction blending. In both strategies, the B-spline curve $\mathbf{C}^\alpha$ is fitted to a target coordinate sequence $\mathbf{D}^{\alpha}=(\mathbf{d}^{\alpha}_1, \dots, \mathbf{d}^{\alpha}_{M^{\alpha}_{D}})$ of edge $b_\alpha$, consisting of $M_D^{\alpha}$ ordered coordinates. In the global construction strategy, we set $\mathbf{D}^{\alpha}=\tilde{\mathbf{Q}}^{\alpha}$, where $\tilde{\mathbf{Q}}^{\alpha}$ denotes the coordinate sequence obtained by pruning $\mathbf{Q}^{\alpha}$ to remove coordinates near each junction node (Section~\ref{subsubsec:global-construction}); in the hierarchical construction strategy, we instead set $\mathbf{D}^{\alpha}=\mathbf{Q}^{\alpha}$. For each coordinate $\mathbf{d}^{\alpha}_\kappa$, $\kappa = 1, \dots, M_D^{\alpha}$, we first compute the corresponding B-spline parameter $\bar{u}_\kappa$ by chord-length parameterization \cite{piegl_nurbs_1995}:
\begin{equation}
\bar{u}_\kappa \;=\; \frac{\sum_{j=2}^{\kappa} \|\mathbf{d}_j^{\alpha} - \mathbf{d}_{j-1}^{\alpha}\|}{\sum_{j=2}^{M^{\alpha}_D} \|\mathbf{d}_j^{\alpha} - \mathbf{d}_{j-1}^{\alpha}\|}, \qquad \bar{u}_1 = 0, \quad \bar{u}_{M^{\alpha}_D} = 1.
\label{eq:chord_length}
\end{equation} The fitting loss is defined as the mean squared distance between the B-spline curve $\mathbf{C}^{\alpha}$ and the target coordinate sequence $\mathbf{D}^{\alpha}$: \begin{equation}
\mathcal{L}^{\alpha}\big(\mathbf{C}^{\alpha}\big) \;=\; \frac{1}{M^{\alpha}_D} \sum_{\kappa=1}^{M^{\alpha}_D} \big\|\, \mathbf{C}^{\alpha}(\bar{u}_\kappa) - \mathbf{d}_\kappa^{\alpha} \,\big\|^2.
\label{eq:fit_loss}
\end{equation} In both construction strategies, the B-spline control point positions, $\mathbf{P}^\alpha$, are optimized to minimize the loss function $\mathcal{L}^{\alpha}$ using Adam optimization~\cite{kingma2014adam}. Fig.~\ref{fig:skeleton_to_curve} summarizes the complete pipeline from skeleton graph construction to B-spline centerline curve fitting.

\subsubsection{Global construction strategy}
\label{subsubsec:global-construction}
In the global construction strategy, the entire branch graph $\mathcal{G}_B$ is extracted from $\mathcal{G}_S$. Before B-spline curve  fitting, the ordered coordinate sequence $\mathbf{Q}^{\alpha}$ is pruned near each junction node to leave room for subsequent junction blending between adjacent B-spline volumes (Section~\ref{sec:junction_continuity}). At each junction node $\nu_\delta\in\mathcal{V}_B^{\mathrm{junc}}$, we define a pruning sphere of radius $R^{\mathrm{prune}}_{\delta}$ centered at its embedded coordinate $\mathbf{x}_\delta$. $R^{\mathrm{prune}}_{\delta}$ is set as the largest radius of the cross-section of all the branches incident at $\nu_\delta$. Any coordinate of $\mathbf{Q}^{\alpha}$ that lies inside this sphere is removed, and the surviving ordered coordinate sequence forms the target coordinate set $\tilde{\mathbf{Q}}^{\alpha}$ used for B-spline curve fitting. The detailed steps for the pruning procedure are outlined in Algorithm~\ref{alg:prune}.

\begin{figure}[pos=htbp]
\begin{algbox}{Junction pruning procedure.}\label{alg:prune}
\item \textbf{Input:} branch graph $\mathcal{G}_B = (\mathcal{V}_B, \mathcal{E}_B)$, ordered coordinate sequence $\{\mathbf{Q}^{\alpha}\}_{b_{\alpha} \in \mathcal{E}_B}$.
\item For each junction node $\nu_\delta \in \mathcal{V}_B^{\text{junc}}$:\begin{itemize}\setlength{\itemsep}{1pt}
\item Set the radius $R^{\mathrm{prune}}_\delta$ of the pruning sphere centered at the embedded coordinate $\mathbf{x}_\delta$.
\item For each edge $b_{\alpha}$ incident to $\nu_\delta$, remove from $\mathbf{Q}^{\alpha}$ every coordinate $\mathbf{q}^{\alpha}$ with $\lVert \mathbf{q}^{\alpha} - \mathbf{x}_\delta \rVert \leq R^\mathrm{prune}_\delta$.
\end{itemize}
\item \textbf{Output:} the surviving ordered coordinate sequence, denoted by $\{\tilde{\mathbf{Q}}^{\alpha}\}_{b_{\alpha} \in \mathcal{E}_B}$.
\end{algbox}
\end{figure}

For each edge $b_\alpha$, the number of control points for each centerline B-spline curve is assigned based on its branch length, with longer branches assigned more control points. The parameters of the B-spline curve are calculated for the target sequence $\tilde{\mathbf{Q}}^{\alpha}$ using Eq.~\ref{eq:chord_length}. Finally, for each edge $b_\alpha$, the B-spline curve $\mathbf{C}^{\alpha}$ is fitted to $\tilde{\mathbf{Q}}^{\alpha}$ by minimizing the loss $\mathcal{L}^{\alpha}$ with respect to $\mathbf{P}^{\alpha}$ (Eq.~\eqref{eq:fit_loss}). In this construction strategy, an independent B-spline curve is fitted to each branch, so adjacent branch curves meeting at the junction do not share a common tangent direction. Tangential continuity is therefore not guaranteed and is enforced separately during junction blending between the branch volumes (Section~\ref{sec:junction_continuity}). Algorithm~\ref{alg:global} outlines the steps carried out in the global construction strategy.
\begin{figure}[pos=htbp]
\begin{algbox}{Global construction strategy.}\label{alg:global}
\item Extract the complete branch graph $\mathcal{G}_B = (\mathcal{V}_B, \mathcal{E}_B)$ from the skeleton graph $\mathcal{G}_S$.
\item Compute the branch lengths $\{L^{\alpha}\}_{b_{\alpha} \in \mathcal{E}_B}$ and partition them into $N_{\text{bin}}$ equal-count (quantile) bins.
\item For each edge $b_{\alpha} \in \mathcal{E}_B$:\begin{itemize}\setlength{\itemsep}{1pt}
\item Assign $n^{\alpha}$ according to the length-bin index of $L^{\alpha}$, with $n^{\min} \leq n^{\alpha} \leq n^{\max}$. Here, $n^{\min}$ and $n^{\max}$ refer to the minimum and maximum number of control points allowed for any branch, respectively.
\item Apply junction pruning to the ordered coordinate sequence $\mathbf{Q}^{\alpha}$ to obtain  $\tilde{\mathbf{Q}}^{\alpha}$ (Algorithm~\ref{alg:prune}).
\item Assign $\{\bar{u}_\kappa\}$ to the pruned coordinate sequence $\tilde{\mathbf{Q}}^{\alpha}$ using Eq.~\eqref{eq:chord_length}.
\item Perform B-spline curve fitting by optimizing the control point positions $\textbf{P}^{\alpha}$ to minimize Eq.~\eqref{eq:fit_loss} using Adam optimization.
\end{itemize}
\end{algbox}
\end{figure} 

\subsubsection{Hierarchical construction strategy}
\label{subsubsec:hierarchical-construction}
In the hierarchical construction strategy, we construct the complete branch graph recursively, starting from a primary path and progressively attaching child branches according to the proximal-to-distal branching order. A primary path, referred to as the main stem, is identified as the path from the root node to a selected terminal leaf node in $\mathcal{G}_S$. The branch graph $\mathcal{G}_B$ for the main stem is initialized with the root node $\nu_{\mathrm{root}}$, the terminal leaf node $\nu_{\alpha}$, and a single directed edge $b_{\alpha} \in \mathcal{E}_B$ connecting them. By default, the terminal leaf node $\nu_{\alpha}$ is selected as the leaf node at the end of the longest root-to-leaf path; although a user-defined leaf node can also be selected. Let $\mathbf{Q}^{\alpha}$ denote the ordered coordinate sequence along the edge $b_{\alpha}$. The coordinates in $\mathbf{Q}^{\alpha}$ are used as target coordinates for B-spline curve fitting (Eqs.~\eqref{eq:chord_length}-\eqref{eq:fit_loss}). Unlike the global construction strategy, $n^{\alpha}$ for each edge is not set according to branch length-based binning, as not all the branch edges are available at initialization. Instead, $n^{\alpha}$ is set equal to the number of coordinates in $\mathbf{Q}^{\alpha}$. 

Next, the child branches in the first branching order are attached to the main stem $b_{\alpha}$, whose fitted centerline B-spline curve is given as $\mathbf{C}^{\alpha}$. Let $\nu_\delta$ be a junction node along the main stem path at which a path in $\mathcal{G}_S$ branches off, and let $\nu_\gamma$ be the distal node of that path; together, they define the child edge $b_{\gamma}$. The node $\nu_\delta$ corresponds to the parameter $\bar{u}_\delta$ on the fitted B-spline curve $\mathbf{C}^{\alpha}$. At two adjacent parameters on either side of $\bar{u}_\delta$, $\bar{u}_\delta^-$ and $\bar{u}_\delta^+$, the knot multiplicity of the knot vector $\mathcal{U}^{\alpha}$ is raised to $p+1$ by repeated knot insertions. As a result, the B-spline curve $\mathbf{C}^{\alpha}$ is split into three curves defined in the parametric intervals $[0, \bar{u}_\delta^-]$, $[\bar{u}_\delta^-, \bar{u}_\delta^+]$, and $[\bar{u}_\delta^+, 1]$. Because knot insertion is a geometry preserving operation \cite{piegl_nurbs_1995, boehm1985insertion}, the three curves together exactly reproduce the geometry of $\mathbf{C}^{\alpha}$. The B-spline curve defined in the parametric interval $[\bar{u}_\delta^-, \bar{u}_\delta^+]$ is deleted to provide sufficient room for subsequent junction blending. This step serves a role similar to the junction pruning procedure carried out in Section~\ref{subsubsec:global-construction}. At initialization, the edge set in the branch graph $\mathcal{G}_B$ is given as $\mathcal{E}_B=\{b_{\alpha}\}$, and it is updated as follows. A junction node $\nu_\delta$ is added to the main stem $b_{\alpha}$. The edge $b_{\alpha}$ is now defined between the nodes $\nu_\delta$ and $\nu_{\alpha}$, and its centerline B-spline curve is defined on the parameter interval $[\bar{u}_\delta^+, 1]$. A new edge, $b_{\delta}$, is defined between the nodes $\nu_{\mathrm{root}}$ and $\nu_{\delta}$, with its corresponding B-spline curve defined on the parameter interval $[0, \bar{u}_\delta^-]$. A child edge $b_{\gamma}$ is then defined from $\nu_\delta$ to $\nu_\gamma$ in the updated branch graph, with $\mathcal{E}_B=\{b_{\delta}, b_{\alpha}, b_{\gamma}\}$.

\begin{figure}[pos=htbp]
\begin{algbox}{Hierarchical construction strategy.}\label{alg:hierarchical}
\item Identify the main stem in the input skeleton graph $\mathcal{G}_S$ as the path from the root node to a terminal leaf node. The terminal leaf node is selected either by a user-defined strategy or, by default, as the leaf node of the longest root-to-leaf path. Initialize the branch graph $\mathcal{G}_B$ with a single edge $b_{\alpha}$ along this path, with the ordered coordinate sequence along this edge defined as $\mathbf{Q}^{\alpha}$.
\item Evaluate the B-spline curve parameters corresponding to the coordinates in $\mathbf{Q}^{\alpha}$ using Eq.~\eqref{eq:chord_length}, and fit the centerline B-spline curve $\mathbf{C}^{\alpha}$ by optimizing the control points $\mathbf{P}^{\alpha}$ to minimize $\mathcal{L}^{\alpha}$ (Eq.~\eqref{eq:fit_loss}). The number of control points is set equal to the number of coordinates in $\mathbf{Q}^{\alpha}$.
\item Traverse the remaining edges of $\mathcal{G}_S$ in a depth-first manner to generate the complete branch graph, moving from each parent edge to its child edges. For each parent edge $b_{\alpha}$ defined between its proximal node $\nu_{\beta}$ and distal node $\nu_{\alpha}$:
    \begin{itemize}\setlength{\itemsep}{1pt}
        \item Sort the children of $b_{\alpha}$ in proximal-to-distal order, from the root node to the terminal leaf nodes.
        \item For each child edge $b_{\gamma}$, attach it to $b_{\alpha}$ at the junction node $\nu_{\delta}$ through the following steps:
            \begin{itemize}\setlength{\itemsep}{1pt}
                \item[(i)]   Determine the parameter $\bar{u}_\delta$ in $\mathcal{U}^{\alpha}$ and the two adjacent parameters on either side of it, $\bar{u}_\delta^-$ and $\bar{u}_\delta^+$, defined on the parent B-spline curve $\mathbf{C}^{\alpha}$.  
                \item[(ii)]  Insert knots at $\bar{u}_\delta^-$ and $\bar{u}_\delta^+$ by repeated single knot insertion until each knot reaches multiplicity $p+1$, creating three B-spline curves on the intervals $[0, \bar{u}_\delta^-]$, $[\bar{u}_\delta^-, \bar{u}_\delta^+]$, and $[\bar{u}_\delta^+, 1]$.
                \item[(iii)] Remove the B-spline curve between $[\bar{u}_\delta^-, \bar{u}_\delta^+]$ to leave room for junction blending. 
                \item[(iv)] $b_{\alpha}$ is now defined between $\nu_\delta$ and $\nu_\alpha$. A new branch edge, $b_{\delta}$, is defined between $\nu_{\beta}$ and $\nu_\delta$.
                \item[(v)]  The child edge $b_{\gamma}$ is created between the nodes $\nu_\delta$ and $\nu_\gamma$, and B-spline curve fitting is performed to evaluate the centerline curve $\mathbf{C}^{\gamma}$.
                \item[(vi)] The child edge $b_{\gamma}$ is now defined as the parent edge for its children.
            \end{itemize}
    \end{itemize}
\end{algbox}
\end{figure}

The remaining branches are added similarly to the branch graph by carrying out the knot insertion and child branch attachment steps. Each edge with a newly fitted B-spline curve subsequently serves as the parent edge for its descendants, and the sub-paths at the junction nodes along this path are attached as its child edges. The path traversal proceeds in a depth-first manner to generate the complete branch graph. Thus, when a child edge is added, all the child edges of that child are processed first before moving to the next child of the same parent. Each child branch attachment operation adds a junction node, $\nu_\delta$, creates a new edge, $b_{\delta}$, updates the edge, $b_{\alpha}$, and finally adds a child edge, $b_{\gamma}$. Thus, the branch graph $\mathcal{G}_B$ grows recursively from a single main stem edge into the complete branch graph, terminating once every point in the skeleton point cloud is fitted to the centerline B-spline curve. The advantage of the hierarchical construction strategy is that the knot insertion operation subdivides a B-spline curve without changing its geometry. Thus, regardless of how many times the parent B-spline curve has been subdivided to attach child branches, it still maintains its geometry and intrinsic smoothness while being represented as multiple B-spline curve segments compared to the global construction strategy. The hierarchical construction strategy is described in Algorithm~\ref{alg:hierarchical}.

\subsection{B-spline volumetric parameterization for each branch}
\label{sec:frames_and_solids}
\begin{figure}[pos=htbp]
    \centering
    \includegraphics[width=0.75\linewidth]{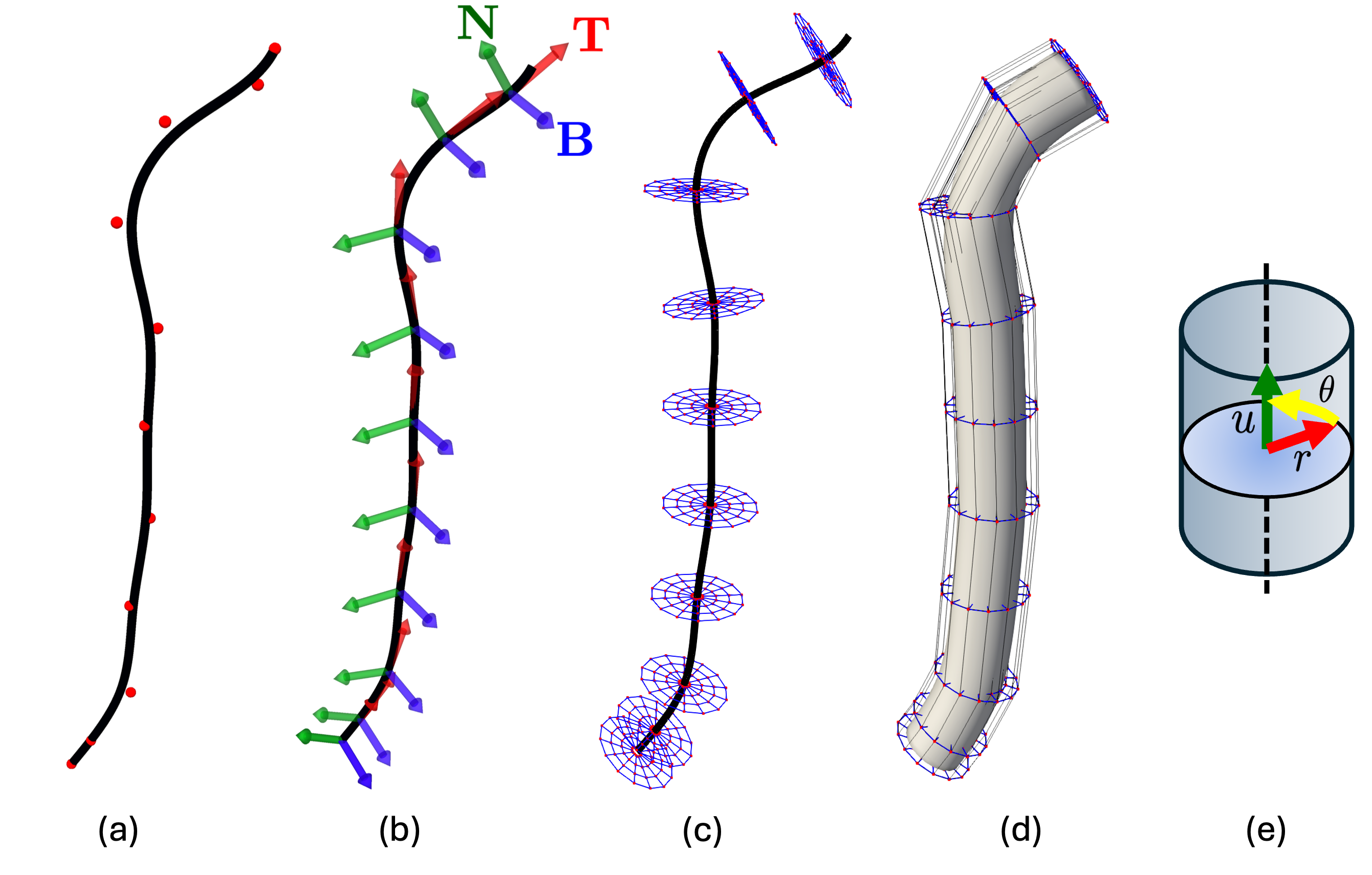}
    \caption{\textbf{B-spline volumetric parametrization of a single branch.} \textbf{(a)} Fitted centerline B-spline curve. \textbf{(b)} Rotation-minimizing frames defined along the centerline curve, with tangent (red), normal (green), and binormal (blue) vectors defined at each Greville abscissa on the curve. \textbf{(c)} Circular cross-sections swept along the centerline curve. \textbf{(d)} Evaluated tensor-product B-spline volume with its corresponding control-point lattice. \textbf{(e)} Schematic showing the cylindrical parameterization of the B-spline volume and the radial, circumferential, and axial parametric directions $(r, \theta, u)$.}
    \label{fig:curve-to-volume}
\end{figure}

Each branch edge $b_\alpha \in \mathcal{E}_{B}$ of the branch graph $\mathcal{G}_B$ is represented by a cylindrical trivariate tensor-product B-spline volume $\mathbf{V}^{\alpha}:[0,1]^3\rightarrow\mathbb{R}^3$ as
\begin{equation}
    \mathbf{V}^\alpha(r, \theta, u) \;=\; \sum_{i=1}^{\ell}\sum_{j=1}^{m}\sum_{k=1}^{n^\alpha} N_{i,s}(r)\, N_{j,q}(\theta)\, N_{k,p}(u)\, \mathbf{P}^\alpha_{i,j,k}, \quad (r, \theta, u)\in[0,1]^3,
    \label{eq:volume}
\end{equation} where $r$, $\theta$, and $u$ are the parametric coordinates, and $s$, $q$, and $p$ are the degrees of the B-spline basis functions in the radial, circumferential, and axial directions, respectively. The B-spline control lattice $\mathbf{P}^\alpha$ is of the size $(\ell \times m \times n^\alpha)$. $\ell$ and $m$ are kept constant for all branches to facilitate better junction blending between the incident volumes. Each basis function $N_{i, s}(r)$, $N_{j, q}(\theta)$, and $N_{k, p}(u)$ is defined on the clamped knot vector $\mathcal{R}$, the unclamped knot vector $\Theta$, and the clamped knot vector $\mathcal{U}^\alpha$, respectively. The circumferential knot vector, $\Theta$, is unclamped in order to define periodic B-splines along the circumferential direction. To construct periodic B-splines, the last $q$ columns are overlapped with the first $q$ columns of the control lattice. Thus, \begin{equation}
    \mathbf{P}^\alpha_{i,\, m - q + w,\, k} \;=\; \mathbf{P}^\alpha_{i,\, w,\, k} \;, \quad w = 1, 2, \dots, q,
   \label{eq:periodicity}
\end{equation} ensures $C^{q-1}$ surface continuity in the circumferential direction \cite{moola2026valvefit}. The tensor-product B-spline volume for each edge $\mathbf{V}^\alpha$ is thus constructed by sweeping the $(r, \theta)$ parameterized cross-sections along the centerline curve $\mathbf{C}^\alpha$ defined in Section~\ref{subsec:centerline-curve-construction} (Fig.~\ref{fig:curve-to-volume}(a)).

A local orthonormal frame $\{\mathbf{T}_k, \mathbf{N}_k, \mathbf{B}_k\}_{k=1}^{n^\alpha}$ is constructed along the centerline curve $\mathbf{C}^\alpha$ at each Greville abscissa $\{\xi_k\}_{k=1}^{n^\alpha}$ (Fig.~\ref{fig:curve-to-volume}(b)), defined on 
$\mathcal{U}^{\alpha}$ \cite{piegl_nurbs_1995}. The unit tangent vector $\mathbf{T}_k$ is tangent to the centerline curve at $\mathbf{C}^{\alpha}(\xi_k)$, whereas the normal and binormal vectors, $\mathbf{N}_k$ and $\mathbf{B}_k$, span a cross-sectional plane that is perpendicular to $\mathbf{T}_k$. The tangent vector is evaluated as $\mathbf{T}_k = \dot{\mathbf{C}}^\alpha(\xi_k) / \|\dot{\mathbf{C}}^\alpha(\xi_k)\|$, where $\dot{\mathbf{C}}^\alpha(\xi_k) =d\mathbf{C}^\alpha(\xi_k) / {du}$. The unit binormal vector is evaluated as the cross-product of the tangent and normal vectors, $\mathbf{B}_k = \mathbf{T}_k \times \mathbf{N}_k$. The first unit normal vector along the centerline curve $\mathbf{N}_1$ is obtained by projecting a fixed global up-vector, $(0, 0, 1)$, onto the plane perpendicular to $\mathbf{T}_1$. If $\mathbf{T}_1$ is nearly parallel to the default up-vector, an alternative up-vector is used. The remaining normals along the centerline curve are propagated using the double-reflection parallel transport method \cite{wang2008rmf}, producing a rotation-minimizing frame by minimizing twist about the tangential direction.

The control points of $\mathbf{V}^\alpha$ are evaluated by sweeping circular cross-sections along the centerline curve $\mathbf{C}^\alpha$ at the Greville abscissae $\{\xi_k\}_{k=1}^{n^\alpha}$. At each $\xi_k$, we define an angular location, $\phi_j$, evaluated via uniform sampling over the circumference of the cross section. Similarly, at each $\xi_k$, a radial location is defined as $\rho_{i}$, evaluated through uniform sampling between the inner and outer radii of the cross section. This is given as
\begin{align}
    \phi_j &= \frac{2\pi (j - 1)}{m - q}, \quad j = 1, \dots, m - q, \\
    \rho_{i} &= h_k \left[ \eta + (1 - \eta) \tfrac{i-1}{\ell-1} \right], \quad i = 1, \dots, \ell,
    \label{eq:angular_radial_sampling}
\end{align} where $h_k$ denotes the outer branch radius at $\xi_k$, obtained by interpolating the outer radius $R_l$ at each skeleton node $\nu_l$ along the centerline curve $\mathbf{C}^\alpha$. The ratio of the inner branch radius to the outer branch radius, $\eta \in (0, 1)$, is fixed for all branches. Each control point in the control lattice $\mathbf{P}^{\alpha}$ is thus given as:
\begin{equation}
\mathbf{P}^{\alpha}_{i,j,k}
=
\mathbf{C}^{\alpha}(\xi_k)
+
\rho_{i}
\left(
\cos(\phi_j)\,\mathbf{N}_k
+
\sin(\phi_j)\,\mathbf{B}_k
\right),
\qquad
\begin{aligned}
&i=1,\ldots,\ell,\\
&j=1,\ldots,m-q,\\
&k=1,\ldots,n^\alpha.
\end{aligned}
\label{eq:cp_construction}
\end{equation}
The $q$ columns in the control-point lattice with circumferential index $j = m - q + 1, \dots, m$ are evaluated according to the periodicity condition in Eq.~\eqref{eq:periodicity}. By sweeping the cross-sections along the centerline and assembling the 3D control point lattice, we can then evaluate the corresponding tensor-product B-spline volume that produces the hollow cylindrical branch geometry shown in Fig.~\ref{fig:curve-to-volume}(c-d).

\subsection{Junction blending for continuity enforcement}
\label{sec:junction_continuity}
As described in Section~\ref{sec:frames_and_solids}, each plant branch is represented by an independent B-spline volume constructed around its own centerline curve. Consequently, at a junction node $\nu_\delta\in\mathcal{V}_B^{\mathrm{junc}}$, the junction-facing boundary cross-sections of the incident branch volumes are generally not coincident. The incident branch volumes must therefore be blended at the junction so that the assembled volumetric mesh contains no gaps and maintains smooth transitions across the junctions. In this section, we describe how these continuity conditions are enforced on the branch control lattices before mesh evaluation. We first describe the $G^1$-continuity requirements for a pair of B-spline curves in Section~\ref{subsubsec:g1-curves} and subsequently describe the extension for a pair of B-spline volumes in Section~\ref{subsubsec:g1-volumes}. We then introduce a cross-section splitting technique for the three incident branch control lattices meeting at bifurcation junctions in Section~\ref{subsubsec:bifurcations}, and finally extend this approach to trifurcation junctions in Section~\ref{subsubsec:trifurcations}. The continuity conditions enforced on the control lattices at each junction are reflected in the evaluated hex mesh. While this approach can be generalized to higher-valence junctions, extensions beyond trifurcations are outside the scope of the present work.

\subsubsection{Junction continuity between two B-spline curves} \label{subsubsec:g1-curves}
\begin{figure}[pos=htbp]
    \centering    \includegraphics[width=0.75\linewidth]{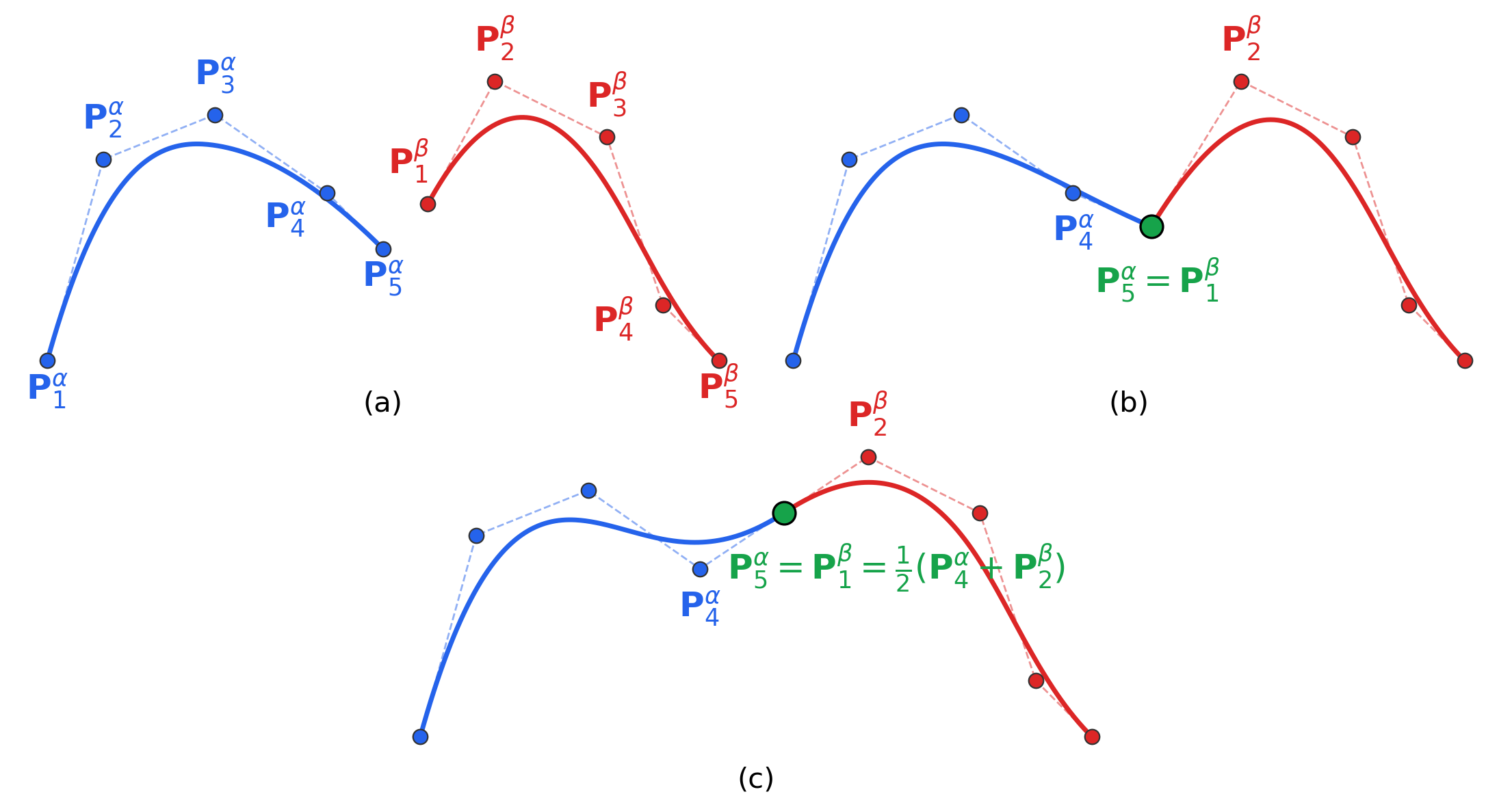}
    \caption{\textbf{$G^1$-continuity enforcement between two cubic B-spline curves.} \textbf{(a)} Before enforcement, the two curves are separated by a gap. \textbf{(b)} After $G^0$ enforcement, the last control point of the first curve and the first control point of the second curve are made coincident, resulting in a shared control point at the junction. \textbf{(c)} After $G^1$ enforcement, the shared control point is placed at $\mathbf{P}^\alpha_5 = \mathbf{P}^\beta_1 = \frac{1}{2}(\mathbf{P}^\alpha_4 + \mathbf{P}^\beta_2)$, making the three unique neighboring control points collinear: $\mathbf{P}^\alpha_4$, $\mathbf{P}^
    \alpha_5 = \mathbf{P}^\beta_1$, and $\mathbf{P}^\beta_2$.}
    \label{fig:g1_continuity_curves}
\end{figure}
Consider two clamped B-spline curves, $\mathbf{C}^{\alpha}$ and $\mathbf{C}^{\beta}$, shown in Fig.~\ref{fig:g1_continuity_curves}(a), with control point sequences $\mathbf{P}^\alpha_1, \dots, \mathbf{P}^\alpha_{n^{\alpha}}$ and $\mathbf{P}^\beta_1, \dots, \mathbf{P}^\beta_{n^{\beta}}$, respectively. Here, $n^{\alpha}$ and $n^{\beta}$ denote the numbers of control points defining $\mathbf{C}^{\alpha}$ and $\mathbf{C}^{\beta}$, respectively. $G^0$-continuity is enforced by setting the terminal control point of $\mathbf{C}^{\alpha}$ equal to the first control point of $\mathbf{C}^{\beta}$. Thus, setting $\mathbf{P}^\alpha_{n^{\alpha}} = \mathbf{P}^\beta_1$ closes the gap between the curves and enforces $G^0$-continuity, as shown in Fig.~\ref{fig:g1_continuity_curves}(b). To achieve $G^1$-continuity, the tangent directions of the two curves must also be aligned at the junction. This can be imposed by making the shared control point and its immediate neighboring control points collinear \cite{piegl_nurbs_1995, zhang2007patient}. We impose both the continuity conditions simultaneously by placing the shared control point at the midpoint of the two adjacent control points: 
\begin{equation}
    \mathbf{P}^\alpha_{n^{\alpha}} = \mathbf{P}^\beta_1 = \tfrac{1}{2}(\mathbf{P}^\alpha_{n^{\alpha}-1} + \mathbf{P}^\beta_2).
    \label{eq:g1_midpoint}
\end{equation} 
This ensures that the control points $\mathbf{P}^\alpha_{n^{\alpha}-1}$, $\mathbf{P}^\alpha_{n^{\alpha}}=\mathbf{P}^\beta_1$, and $\mathbf{P}^\beta_2$ are collinear (Fig.~\ref{fig:g1_continuity_curves}(c)), thereby providing a common tangent direction at the junction.

\subsubsection{Junction continuity between two B-spline volumes}
\label{subsubsec:g1-volumes}
\begin{figure}[pos=htbp]
    \centering
    \includegraphics[width=0.8\linewidth]{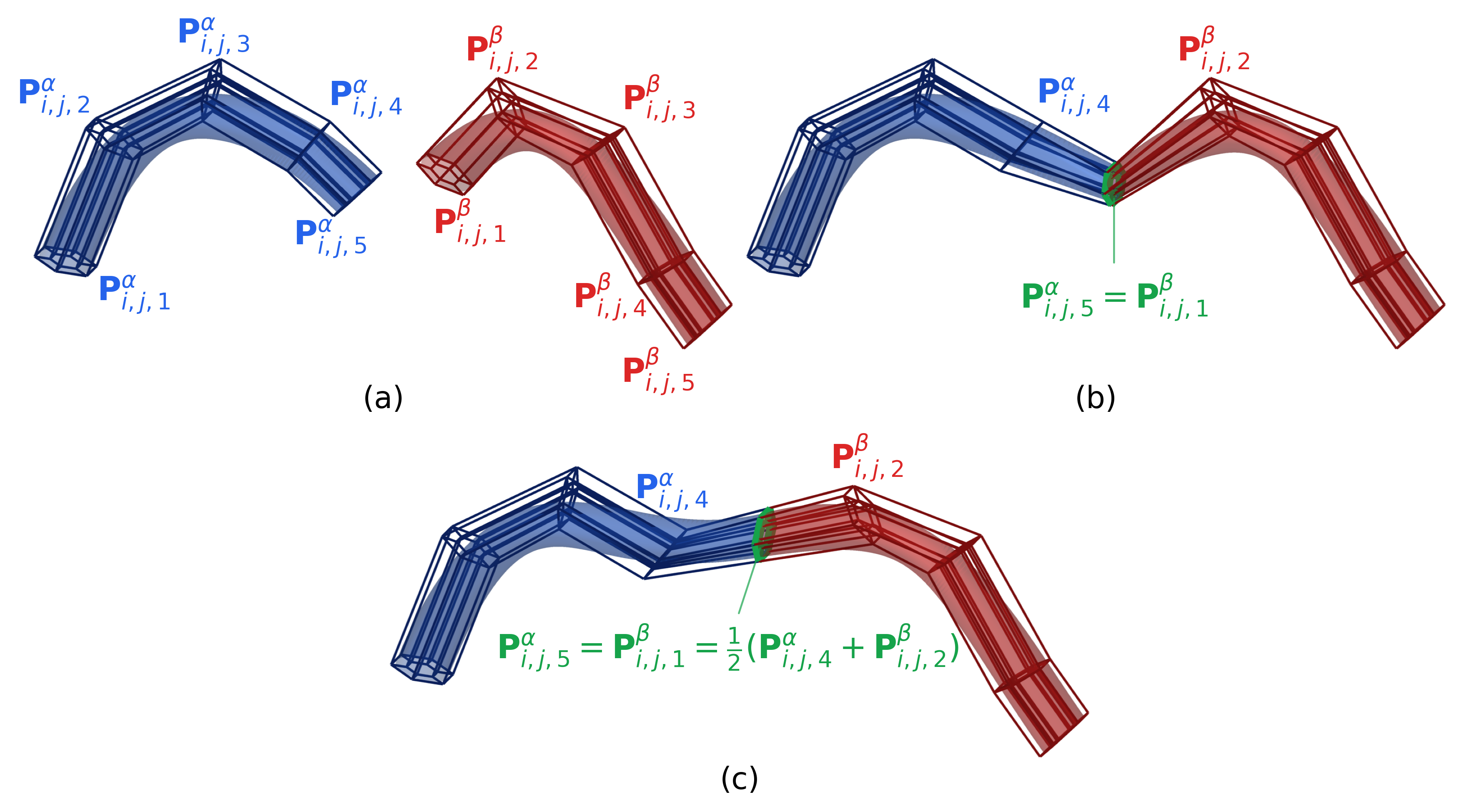}
    \caption{\textbf{$G^1$-continuity enforcement between two cubic trivariate tensor-product B-spline volumes.} The control lattices of the two B-spline volumes are denoted by $\mathbf{P}^{\alpha}$ and $\mathbf{P}^{\beta}$. A control point in $\mathbf{P}^{\alpha}$ is denoted by $\mathbf{P}^{\alpha}_{i,j,k}$, where $i$, $j$, and $k$ are the indices in the radial, circumferential, and axial directions, respectively. \textbf{(a)} Before enforcement, the two volumes are separated by a gap. \textbf{(b)} After $G^0$ enforcement, the last control point in each axial control column of the first B-spline volume and the first control point in each axial control column of the second B-spline volume are made coincident, creating a shared boundary control cross-section, shown in green. \textbf{(c)} After $G^1$ enforcement, each control point on the shared boundary control cross-section is placed according to $\mathbf{P}^{\alpha}_{i,j,5} = \mathbf{P}^{\beta}_{i,j,1} = \frac{1}{2}(\mathbf{P}^{\alpha}_{i,j,4} + \mathbf{P}^{\beta}_{i,j,2})$ for all $(i,j)$, so that the corresponding control points across the shared boundary cross section in the axial direction are collinear.}
    \label{fig:g1_continuity_volumes}
\end{figure}
The $G^1$-continuity enforcement for curves in Eq.~\eqref{eq:g1_midpoint} extends naturally to a pair of adjacent trivariate tensor-product B-spline volumes. Consider two B-spline volumes, $\mathbf{V}^{\alpha}$ and $\mathbf{V}^{\beta}$, defined by the 3D control lattices $\mathbf{P}^\alpha$ and $\mathbf{P}^\beta$, respectively. For each fixed radial index, $i$, and circumferential index, $j$, the control points along the axial direction form an axial control column denoted by $\mathbf{P}^{\alpha}_{i,j,:}$ or $\mathbf{P}^{\beta}_{i,j,:}$, as shown in Fig.~\ref{fig:g1_continuity_volumes}. The two B-spline volumes are joined by applying Eq.~\eqref{eq:g1_midpoint} column-wise to the axial control points for every fixed pair $(i, j)$. Specifically, $G^0$-continuity is enforced by making the terminal control point of each axial column in the control lattice of $\mathbf{V}^{\alpha}$ equal to the first control point of each corresponding axial column in the control lattice of $\mathbf{V}^{\beta}$. Thus, $\mathbf{P}^{\alpha}_{i,j,n^{\alpha}} = \mathbf{P}^{\beta}_{i,j,1}$ for all $(i,j)$. This creates a shared boundary control cross-section between the two B-spline volumes, as shown in Fig.~\ref{fig:g1_continuity_volumes}(b). To enforce $G^1$-continuity, each control point on the shared boundary control cross-section is placed at the midpoint of its two neighboring axial control points, $\mathbf{P}^\alpha_{i,j,n^{\alpha}-1}$ and $\mathbf{P}^\beta_{i,j,2}$: \begin{equation}
    \mathbf{P}^\alpha_{i,j,n^{\alpha}} = \mathbf{P}^\beta_{i,j,1}
    = \tfrac{1}{2}\bigl(\mathbf{P}^\alpha_{i,j,n^{\alpha}-1} +\mathbf{P}^\beta_{i,j,2}\bigr),
    \qquad \forall\,(i,j).
    \label{eq:g1_volume_midpoint}
\end{equation}
Thus, three unique neighboring control points in each axial column become collinear, thereby providing a common axial tangent direction across the shared boundary between the two B-spline volumes.

\subsubsection{Junction continuity at bifurcations}
\label{subsubsec:bifurcations}
\begin{figure}[pos=htbp]
    \centering
\includegraphics[width=0.95\linewidth]{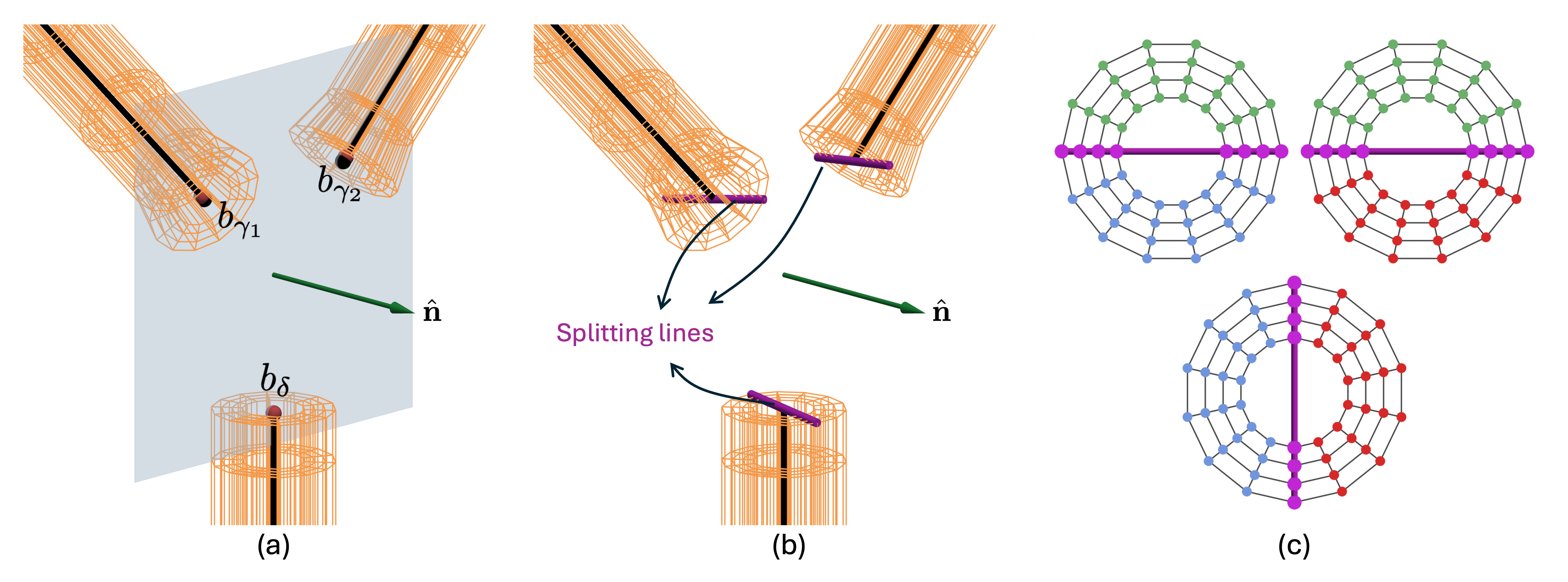}
    \caption{\textbf{Splitting line construction at a bifurcation junction.} \textbf{(a)} The junction plane, shown in gray, is defined by the junction-facing endpoints of the parent and child centerline curves. The unit vector perpendicular to the junction plane $\hat{\mathbf{n}}$ is shown. \textbf{(b)} On the junction-facing control-point cross-section of each incident branch, the radial line that is most closely aligned with the direction of $\hat{\mathbf{n}}$ is selected as the splitting line (shown in magenta). \textbf{(c)} The splitting line is the radial diameter that partitions each control lattice cross-section into two halves. Each color indicates the pair of control-lattice axial columns associated with pairwise continuity enforcement between the branch edges ($b_\delta$--$b_{\gamma_1}$ (blue), $b_\delta$--$b_{\gamma_2}$ (red), $b_{\gamma_1}$--$b_{\gamma_2}$ (green)).}
    \label{fig:junction-normal}
\end{figure}

The pairwise continuity enforcement between two B-spline volumes described in Section \ref{subsubsec:g1-volumes} cannot be applied directly to bifurcations because three B-spline volumes meet at the junction. At a bifurcation junction node $\nu_\delta \in \mathcal{V}_B^{\mathrm{junc}}$ in $\mathcal{G}_B$, one parent edge $b_\delta$ meets two child edges: $b_{\gamma_1}$ and $b_{\gamma_2}$. Let $a \in \{\delta, \gamma_1, \gamma_2\}$ index the three branch edges incident at $\nu_\delta$. For each incident edge $b_a$, let $\mathbf{C}^a$ denote its centerline curve, $\mathbf{V}^a$ denote its tensor-product B-spline volume, and $\mathbf{P}^a$ denote its corresponding control lattice. A control point in this lattice is denoted by $\mathbf{P}^a_{i,j,k}$, where $i$, $j$, and $k$ are the indices along the radial, circumferential, and axial directions, respectively. To enforce continuity at the bifurcation, the junction-facing control cross-section of each incident B-spline volume is first partitioned into two circumferential halves, as shown in Fig.~\ref{fig:junction-normal}. A \emph{splitting line} is defined on each junction-facing control cross-section to partition the axial control columns into two circumferential halves. Corresponding axial control columns from paired circumferential halves are then blended using  Eq.~\eqref{eq:g1_volume_midpoint}. Because all incident control lattices have the same radial and circumferential resolution, corresponding circumferential halves can be paired across the three control lattices.

We define a \emph{junction plane} that passes through the endpoints of the three centerline curves near the junction, as shown in Fig.~\ref{fig:junction-normal}(a). Let $\mathbf{p}^{\delta} = \mathbf{C}^{\delta}(1)$, $\mathbf{p}^{\gamma_{1}} = \mathbf{C}^{\gamma_{1}}(0)$, and $\mathbf{p}^{\gamma_{2}} = \mathbf{C}^{\gamma_{2}}(0)$ denote the junction-side endpoints of the parent and child curves, respectively. A unit normal vector to the junction plane, $\hat{\mathbf{n}}$,  is given as 
\begin{equation}
    \hat{\mathbf{n}} \;=\; \frac{(\mathbf{p}^{\gamma_1} - \mathbf{p}^{\delta}) \times (\mathbf{p}^{\gamma_2} - \mathbf{p}^{\delta})}{\|(\mathbf{p}^{\gamma_1} - \mathbf{p}^{\delta}) \times (\mathbf{p}^{\gamma_2} - \mathbf{p}^{\delta})\|}.
    \label{eq:junction_vector}
\end{equation} This unit normal vector provides a common reference direction for selecting the splitting line on each junction-facing control-point cross-section. For each incident branch, a control cross-section at a given axial index is assigned a depth index $d$ based on its distance from the junction. Here, $d=0$ denotes the junction-facing cross-section on each control lattice. For each incident edge $a$, the axial index of each control point on the cross-section at depth-$d$ is 
\begin{equation}
    k^a_d =
    \begin{cases}
        n^a - d, & \text{if } a=\delta \\
        1 + d,   & \text{if } a \in \{\gamma_1, \gamma_2\},
    \end{cases}
    \label{eq:axial-depth-index}
\end{equation} where $n^a$ is the number of control points in the axial direction for the control lattice $\mathbf{P}^{a}$. On the junction-facing control cross-section, with axial index $k^{a}_{0}$, we define a radial line extending from $\mathbf{P}^{a}_{1,j,k^{a}_{0}}$ to $\mathbf{P}^{a}_{\ell,j,k^{a}_{0}}$. The splitting line is chosen as the radial diameter whose direction is most closely aligned with the normal vector $\hat{\mathbf{n}}$. The corresponding circumferential index $j^a_\mathrm{split}$ on the splitting line is thus given as: 
\begin{equation}
    j^a_\mathrm{split} = \arg\max_{j\in\{1, \ldots, m-q\}}
    \left|\,
    \left(\frac{\mathbf{P}^{a}_{\ell,j, k^a_0} - \mathbf{P}^{a}_{1,j, k^a_0}}{\left\| \mathbf{P}^{a}_{\ell,j, k^a_0} - \mathbf{P}^{a}_{1,j, k^a_0} \right\|}\right)\cdot\hat{\mathbf{n}}\,\right|.
    \label{eq:splitting-index}
\end{equation} The selected radial diameter partitions the control-lattice cross-section into two halves. Fig.~\ref{fig:junction-normal}(c) shows the cross-sections partitioned by the splitting line on each control lattice cross-section, where matching colors indicate the corresponding pairs of axial columns used for pairwise continuity enforcement: $b_{\delta}-b_{\gamma_1}$ (blue), $b_{\delta}-b_{\gamma_2}$ (red), and $b_{\gamma_1}-b_{\gamma_2}$ (green). For each paired set of axial columns, the continuity rule in Eq.~\eqref{eq:g1_volume_midpoint} is applied column-wise. The control points on the depth $d=0$ cross-section are placed at the midpoint of their corresponding depth $d=1$ neighboring control points, thereby enforcing $G^{1}$-continuity across that interface. 

The splitting line itself is treated as a common three-way control-lattice interface among the three incident branches at the junction. For each radial index $i$ on the splitting line, we compute the centroid of the three adjacent depth $d=1$ control points as
\begin{equation}
    \bar{\mathbf{P}}_i \;=\; \frac{1}{3} \sum_{a \in \{\delta, \gamma_1, \gamma_2\}}
        \mathbf{P}^{a}_{i, j^{a}_{\text{split}}, k^a_1}.
    \label{eq:junction-average}
\end{equation} The control points on the splitting line of all three incident branches are then assigned to this centroid as
\begin{equation}
    \mathbf{P}^{a}_{i, j^{a}_{\text{split}}, k^a_0} = \bar{\mathbf{P}}_i,
    \qquad \forall\, a \in \{\delta, \gamma_1, \gamma_2\}.
    \label{eq:junction-assign}
\end{equation} This operation assigns a common column of control points to the shared splitting line among the parent and child branch control lattices while using the adjacent depth $d=1$ control columns to define the local tangent alignment at the junction. 

\subsubsection{Junction continuity at trifurcations}
\label{subsubsec:trifurcations}
\begin{figure}[pos=htbp]
    \centering
    \includegraphics[width=0.95\linewidth]{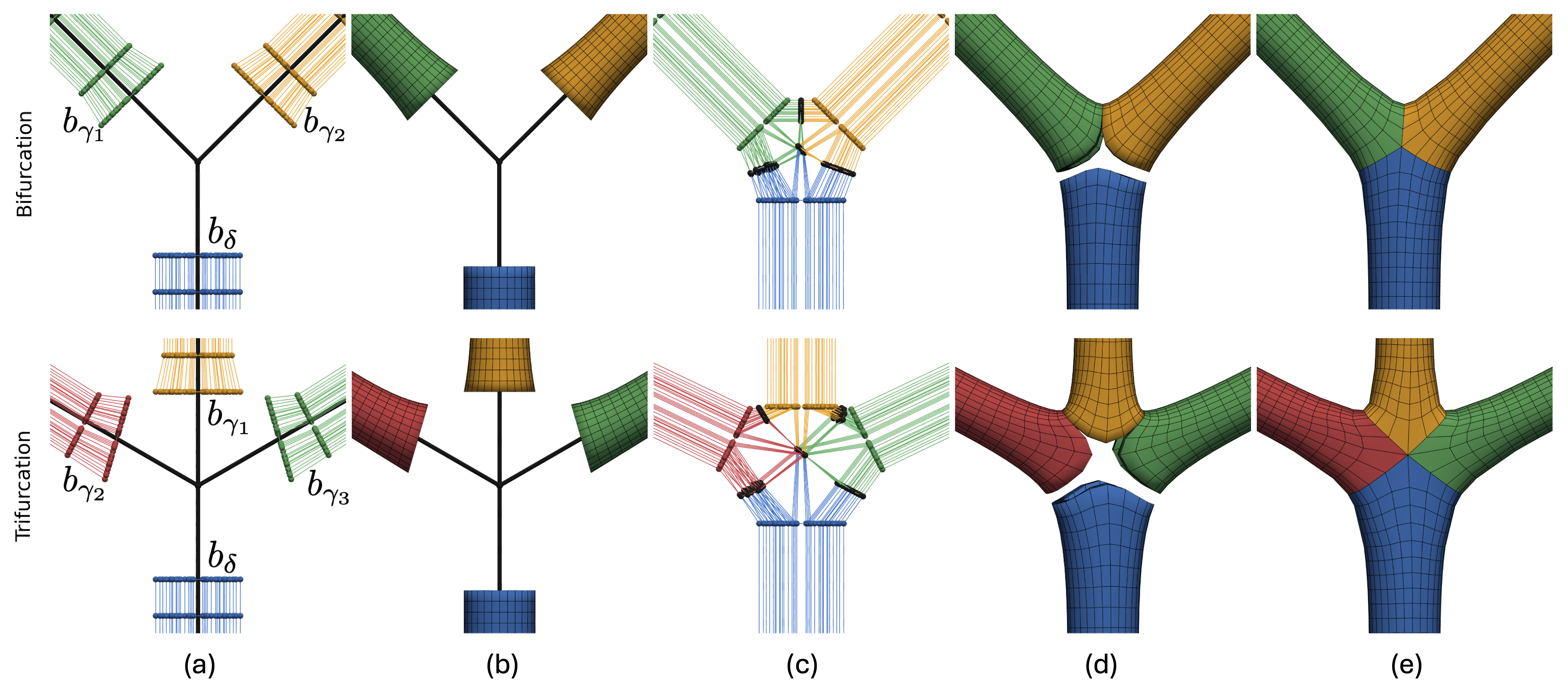}
    \caption{\textbf{Volumetric hex mesh construction at a bifurcation junction (top row) and a trifurcation junction (bottom row).} The parent B-spline volume is shown in blue. For the bifurcation junction, the two child B-spline volumes are shown in green and yellow; for the trifurcation junction, the three child B-spline volumes are shown in red, yellow, and green. \textbf{(a)} Control lattices of the parent and child B-spline volumes before junction blending. \textbf{(b)} Evaluated B-spline branch volumes before junction blending, showing gaps between independently constructed incident volumes near the junction. \textbf{(c)} Control lattices after junction blending, where the control point positions on the junction-facing cross-sections are modified to enforce continuity between incident branches. \textbf{(d)} Evaluated B-spline volumes after junction blending, showing a small residual gap along the splitting-line interface. \textbf{(e)} Smooth, conforming hex mesh obtained after residual gap closure by averaging corresponding adjacent mesh vertices.}
    \label{fig:blending-stages}
\end{figure}
The enforcement of junction continuity at bifurcations, as described in Section~\ref{subsubsec:bifurcations}, extends naturally to trifurcation junctions. At a trifurcation junction node $\nu_\delta \in \mathcal{V}_B^{\mathrm{junc}}$ in $\mathcal{G}_B$, one parent edge $b_\delta$ meets three child edges $b_{\gamma_1}$, $b_{\gamma_2}$, and $b_{\gamma_3}$. Let $a \in \{\delta, \gamma_1, \gamma_2, \gamma_3\}$ index the four branch edges incident at $\nu_\delta$. For each incident edge $b_a$, let $\mathbf{C}^a$ denote its B-spline centerline curve, $\mathbf{V}^a$ denote its B-spline volume, and $\mathbf{P}^a$ denote its corresponding control lattice. As in the case of bifurcation junctions, the junction-facing cross-section of each incident control lattice is split into two halves by selecting a radial diameter, also known as the splitting line. The circumferential index of the splitting line, $j_{\mathrm{split}}^a$, is determined using Eq.~\eqref{eq:splitting-index}. The resulting partitioned control-lattice cross-sections are paired across the four incident edges as follows: $b_{\delta}-b_{\gamma_2}$, $b_{\gamma_2}-b_{\gamma_1}$, $b_{\gamma_1}-b_{\gamma_3}$, and $b_{\gamma_3}-b_{\delta}$, as shown in Fig.~\ref{fig:blending-stages}. For each paired set of axial columns in the control lattices, continuity is enforced using Eq.~\eqref{eq:g1_volume_midpoint}. 

The continuity enforcement procedure at the trifurcation junction differs from that at the bifurcation junction in two ways: the construction of the unit normal vector $\hat{\mathbf{n}}$ and the treatment of the shared splitting-line interface. At a bifurcation junction, the three junction-side endpoints of the centerline curves define a unique junction plane. However, at a trifurcation junction, the endpoints $\mathbf{p}^a$ of the four centerline curves $\mathbf{C}^a$ are not always coplanar. Therefore, $\hat{\mathbf{n}}$ is computed as the unit normal vector to the best-fit plane passing through these four endpoints. Let
\begin{equation}
\mathbf{p}^{\delta} = \mathbf{C}^{\delta}(1),
\qquad
\mathbf{p}^{\gamma_r} = \mathbf{C}^{\gamma_r}(0),
\quad r=1,2,3,
\end{equation} denote the junction-side endpoints of the parent and child centerline curves. The centroid of these four endpoints is $\bar{\mathbf{p}} = \frac{1}{4} \sum_{a \in \{\delta, \gamma_1, \gamma_2, \gamma_3\}} \mathbf{p}^{a}$. We define a $4 \times 3$ matrix $\mathbf{E}$ whose rows are $(\mathbf{p}^{a} - \bar{\mathbf{p}})^{\top}$ for $a \in \{\delta, \gamma_1, \gamma_2, \gamma_3\}$. Applying singular value decomposition (SVD) to $\mathbf{E}$, we obtain $\hat{\mathbf{n}}$ as the right singular vector corresponding to the smallest singular value. The circumferential index on the splitting line $j^a_\mathrm{split}$ is then determined for each control-lattice cross-section using Eq.~\eqref{eq:splitting-index}.

At a trifurcation junction, the splitting line is treated as a common four-way control-lattice interface among the incident B-spline volumes. For each radial index $i$ on the splitting line, we compute the centroid of the four adjacent depth $d=1$ control points as 
\begin{equation}
    \bar{\mathbf{P}}_i = \frac{1}{4}\sum_{a\in\{\delta,\gamma_1,\gamma_2,\gamma_3\}}\mathbf{P}^{a}_{i,\,j^{a}_\mathrm{split},\,k^{a}_1}. \label{eq:centroid}
\end{equation} The control points on the shared splitting line of all four incident branches are then assigned to this centroid: 
\begin{equation}
    \mathbf{P}^a_{i, j^a_\mathrm{split}, k^a_0} = \bar{\mathbf{P}}_i, \quad \forall a \in \{\delta, \gamma_1, \gamma_2, \gamma_3\}.
\label{eq:trifurcation-assign}
\end{equation} However, the four adjacent depth $d=1$ splitting-line control points, $\mathbf{P}^{a}_{i,j^a_\mathrm{split},k^a_1}$, where $a \in \{\delta, \gamma_1, \gamma_2, \gamma_3\}$, need not be coplanar. For each radial index $i$, we construct a $4 \times 3$ centered matrix whose rows are the offsets of these four points from their centroid:
\begin{equation}
    \mathbf{M}_i = \begin{pmatrix}
        \mathbf{P}^\delta_{i, j^\delta_\mathrm{split}, k^\delta_1} - \bar{\mathbf{P}}_i \\
        \mathbf{P}^{\gamma_1}_{i,\,j^{\gamma_1}_\mathrm{split},\,k^{\gamma_1}_1} - \bar{\mathbf{P}}_i \\ 
        \mathbf{P}^{\gamma_2}_{i,\,j^{\gamma_2}_\mathrm{split},\,k^{\gamma_2}_1} - \bar{\mathbf{P}}_i \\ 
        \mathbf{P}^{\gamma_3}_{i,\,j^{\gamma_3}_\mathrm{split},\,k^{\gamma_3}_1} - \bar{\mathbf{P}}_i
    \end{pmatrix}
    \label{eq:adjacent_points_svd}
\end{equation} Applying SVD to $\mathbf{M}_i$, we obtain the local unit normal vector $\hat{\mathbf{m}}_i$ as the right singular vector corresponding to the smallest singular value. The four adjacent depth $d=1$ control points are then projected onto this best-fit plane. We first evaluate $\tilde{\mathbf{P}}^{a}_{i}$ as
\begin{equation}
    \tilde{\mathbf{P}}^{a}_{i} = \mathbf{P}^{a}_{i,\,j^a_\mathrm{split},\,k^a_1} - \Bigl[\hat{\mathbf{m}}_i \cdot \Bigl(\mathbf{P}^{a}_{i,\,j^a_\mathrm{split},\,k^a_1} - \bar{\mathbf{P}}_i\Bigr)\Bigr]\hat{\mathbf{m}}_i, \qquad a \in \{\delta, \gamma_1, \gamma_2, \gamma_3\}.
    \label{eq:assignment}
\end{equation} We then replace the original depth $d=1$ control point $\mathbf{P}^{a}_{i,\,j^a_\mathrm{split},\,k^a_1}$ with its projected point $\mathbf{P}^a_{i, j^a_\mathrm{split}, k^a_1} = \tilde{\mathbf{P}}^{a}_{i}$. This coplanarity condition serves as the multi-way analog of the collinearity condition in the pairwise case (Eq. \ref{eq:g1_volume_midpoint}), and enables common tangent-plane alignment across the four incident B-spline volumes at the junction. Eqs.~\eqref{eq:centroid}--\eqref{eq:assignment} assign a common column of control points to the shared splitting line of the parent and child control lattices while enforcing coplanarity with the adjacent depth $d=1$ control points that determine the local tangent alignment at the junction interface.

\subsubsection{Hexahedral mesh generation}
To generate the hex mesh vertices, each B-spline volume $\mathbf{V}^\alpha$ is evaluated on a structured 3D parametric grid following Eq.~\eqref{eq:volume}. The parametric grid is of the size $S_r \times S_\theta \times S^\alpha_u$, where $S_r$, $S_\theta$, and $S^\alpha_u$ denote the number of parametric coordinates along the radial, circumferential, and axial directions, respectively. The B-spline volume $\mathbf{V}^\alpha$ of each branch edge is then evaluated at these parametric coordinates to generate a structured grid of mesh vertices. By construction, the control points in the control lattice of each B-spline volume are obtained by uniform sampling along the radial and circumferential directions (Eqs.~\eqref{eq:angular_radial_sampling}-\eqref{eq:cp_construction}). Therefore, the uniform sampling of radial and circumferential parametric coordinates results in uniformly placed mesh vertices in each branch cross-section. $S_r$ and $S_\theta$ thus remain constant for all the branch edges, which is  important for enforcing junction continuity. However, along the axial direction, the control points are not uniformly distributed across all the branch edges. Therefore, the parametric coordinates in the axial direction are chosen so that mesh vertices have approximately equal spacing along the branch centerline curve. The number of parametric coordinates along the axial direction $S^\alpha_u$ varies according to the branch length $L^\alpha$. For a fixed spacing between the mesh vertices along the axial direction, $\Delta_u$, we calculate $S^\alpha_u$ as
\begin{equation}
S^\alpha_u = \max\left(S_{\min}, \left\lceil \frac{L^\alpha}{\Delta_u} \right\rceil\right),
\end{equation}
where $S_{\min}$ is a prescribed lower bound for the number of axial parametric coordinates, ensuring that short branches still contain a minimum number of mesh vertices along the axial direction. The control lattices, after enforcing junction blending for bifurcations and trifurcations, are shown in Fig.~\ref{fig:blending-stages}(c), and the corresponding evaluated B-spline volumes are shown in Fig.~\ref{fig:blending-stages}(d). In the circumferential direction, periodic B-splines defined on an unclamped knot vector are non-interpolatory at the control points along the shared splitting-line interface between the incident branches. Therefore, a small residual gap may remain between the evaluated B-spline volumes along the splitting-line interface. This gap is closed during hex mesh construction by averaging corresponding adjacent mesh vertices, producing a conforming hex mesh across each junction, as shown in Fig.~\ref{fig:blending-stages}(e).

\subsection{Mesh quality study at junctions}
We evaluate the performance  of the junction continuity enforcement by varying the angles between the incident branches for bifurcation and trifurcation junctions. For each generated hex mesh, we report the element quality using the scaled Jacobian metric. For an element $e$, let $\mathbf{t}_{A,1}$, $\mathbf{t}_{A,2}$, and $\mathbf{t}_{A,3}$ denote the three edge vectors meeting at a corner $A$, ordered consistently with the right-handed reference hexahedron. Let $L_{A,1}, L_{A,2}, L_{A,3}$ denote the corresponding edge lengths. The Jacobian determinant at the corner $A$ is the signed volume of the parallelepiped spanned by these edge vectors, $J_A = \det[\mathbf{t}_{A,1}, \mathbf{t}_{A,2}, \mathbf{t}_{A,3}]$. The scaled Jacobian at a corner $A$ of the element is evaluated by normalizing $J_A$ by the product of the three incident edge vectors. The scaled Jacobian of the element $e$ is thus given as
\begin{equation}
q_{e}
=
\min_{A=1,\ldots,8}
\frac{J_A}{L_{A,1}L_{A,2}L_{A,3}}.
\label{eq:min_scaled_jacobian}
\end{equation}
This metric satisfies $|q_{e}| \le 1$, with $q_{e} = 1$ corresponding to an orthogonal, positively oriented element. $q_{e} \le 0$ indicates that at least one corner of the element is degenerate or inverted.

\begin{figure}[pos=htbp]
    \centering
    \setlength{\lineskip}{0pt}

    \begin{subfigure}[b]{0.32\linewidth}
        \includegraphics[width=\linewidth]{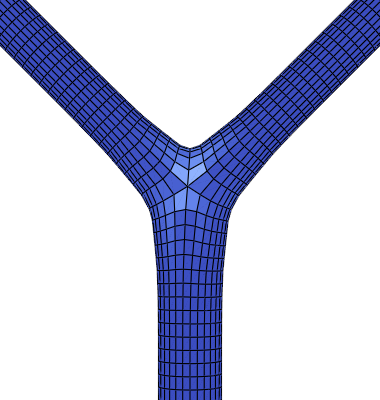}
        \caption{($135\degree$, $90\degree$, $135\degree$)}
    \end{subfigure}%
    \begin{subfigure}[b]{0.32\linewidth}
        \includegraphics[width=\linewidth]{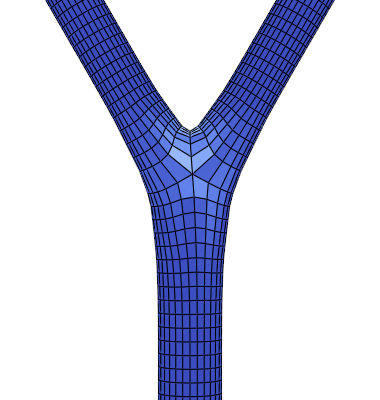}
        \caption{($150\degree$, $60\degree$, $150\degree$)}
    \end{subfigure}%
    \begin{subfigure}[b]{0.32\linewidth}
        \includegraphics[width=\linewidth]{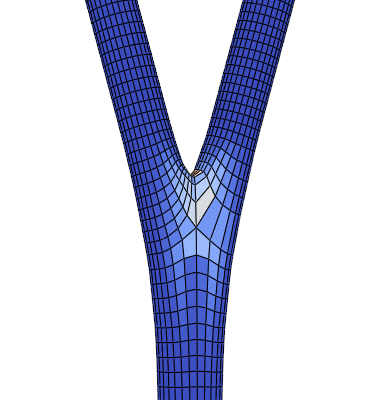}
        \caption{($165\degree$, $30\degree$, $165\degree$)}
    \end{subfigure}

    \begin{subfigure}[b]{0.32\linewidth}
        \includegraphics[width=\linewidth]{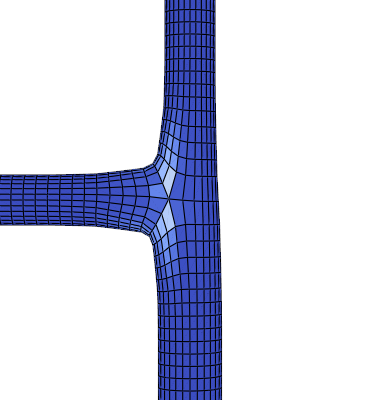}
        \caption{($90\degree$, $90\degree$, $180\degree$)}
    \end{subfigure}%
    \begin{subfigure}[b]{0.32\linewidth}
        \includegraphics[width=\linewidth]{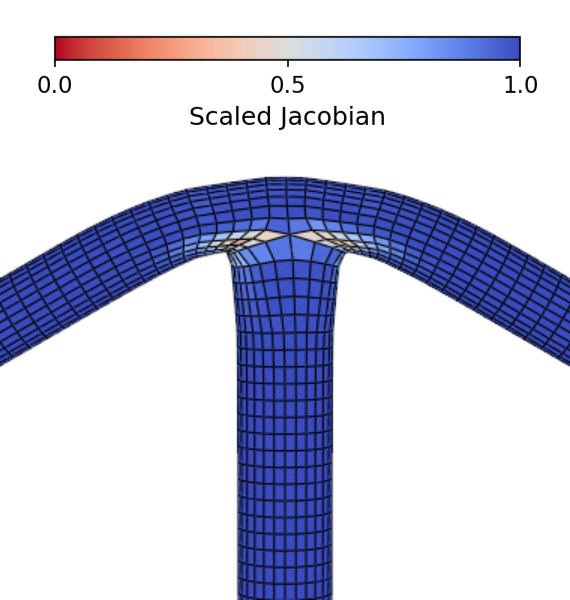}
        \caption{($60\degree$, $240\degree$, $60\degree$)}
    \end{subfigure}%
    \begin{subfigure}[b]{0.32\linewidth}
        \includegraphics[width=\linewidth]{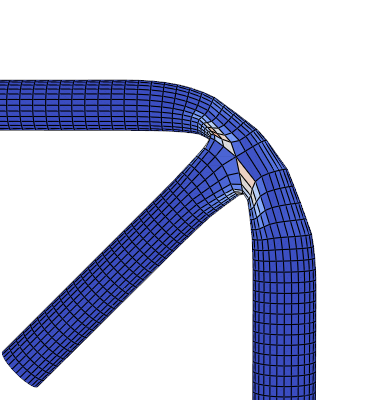}
        \caption{($45\degree$, $45\degree$, $270\degree$)}
    \end{subfigure}
    \caption{\textbf{Angle parametric study for bifurcation junction configurations.} For each configuration, the three listed angles denote the angles, in degrees, between consecutive incident branches around the junction. Each generated mesh is 
colored according to the element-wise scaled Jacobian value.}
    \label{fig:angles_bifurcation}
\end{figure}

\begin{figure}[pos=htbp]
    \centering
    \setlength{\lineskip}{0pt}

    \begin{subfigure}[b]{0.32\linewidth}
        \includegraphics[width=\linewidth]{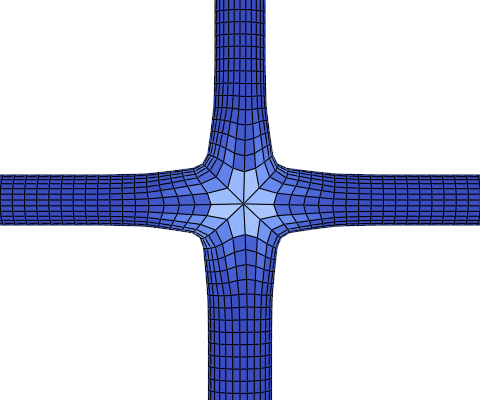}
        \caption{($90\degree$, $90\degree$, $90\degree$, $90\degree$)}
    \end{subfigure}%
    \begin{subfigure}[b]{0.32\linewidth}
        \includegraphics[width=\linewidth]{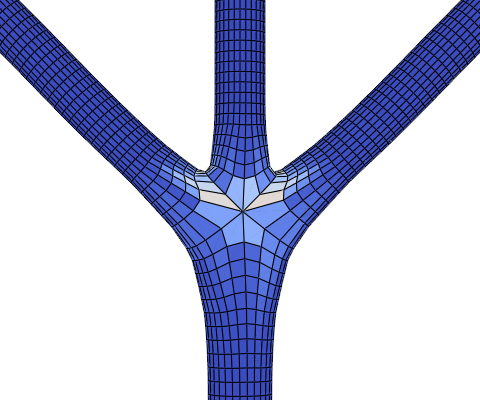}
        \caption{($135\degree$, $45\degree$, $45\degree$, $135\degree$)}
    \end{subfigure}%
    \begin{subfigure}[b]{0.32\linewidth}
        \includegraphics[width=\linewidth]{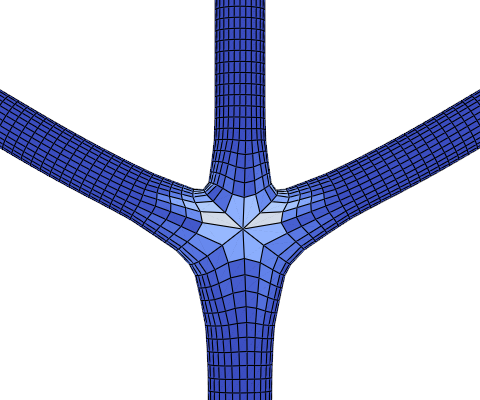}
        \caption{($120\degree$, $60\degree$, $60\degree$, $120\degree$)}
    \end{subfigure}

    \begin{subfigure}[b]{0.32\linewidth}
        \includegraphics[width=\linewidth]{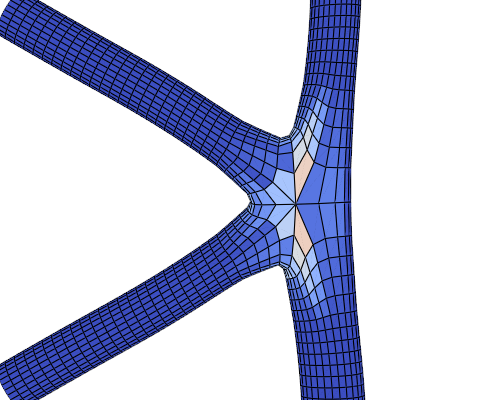}
        \caption{($60\degree$, $60\degree$, $60\degree$, $180\degree$)}
    \end{subfigure}%
    \begin{subfigure}[b]{0.32\linewidth}
        \includegraphics[width=\linewidth]{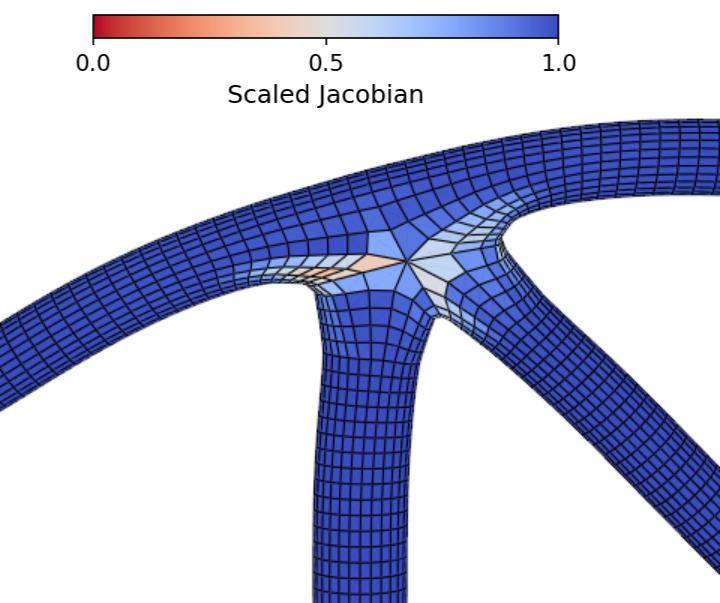}
        \caption{($60\degree$, $210\degree$, $45\degree$, $45\degree$)}
    \end{subfigure}%
    \begin{subfigure}[b]{0.32\linewidth}
        \includegraphics[width=\linewidth]{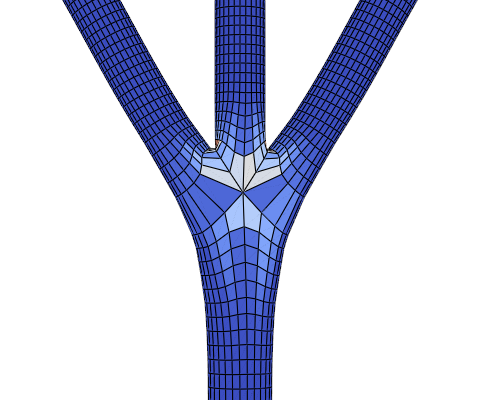}
        \caption{($150\degree$, $30\degree$, $30\degree$, $150\degree$)}
    \end{subfigure}

    \caption{\textbf{Angle parametric study for trifurcation junction configurations.} For each configuration, the four listed angles denote the angles, in degrees, between consecutive incident branches around the junction. Each generated mesh is 
colored according to the element-wise scaled Jacobian value.}
    \label{fig:angles_trifurcation}
\end{figure}

Fig.~\ref{fig:angles_bifurcation} shows different bifurcation junction configurations generated by varying the angle between the three incident branches. For each configuration, the angle between each branch is set to be between $30\degree$ and $270\degree$. For all the bifurcation junction configurations, we report positive scaled Jacobian values, as seen in Table~\ref{tab:angle_quality}. In addition, the $10^\text{th}$ percentile of the element quality ($P_{10}$) satisfies $P_{10} \ge 0.925$, indicating that at least $90\%$ of the mesh elements in each configuration have $q_{e} \ge 0.925$. Similarly, we perform the angle parametric study for trifurcation junctions, as shown in Fig.~\ref{fig:angles_trifurcation}, varying the angles between the four incident branches at the junction. For all the configurations, we report good mesh quality, where all the elements are non-inverted ($q_{e}>0$). We maintain $P_{10} \ge 0.905$ for trifurcation junctions, indicating that at least $90\%$ of the mesh elements in each configuration have $q_{e} \ge 0.905$. In all the junction configurations, the lowest $q_{e}$ value is observed near the junction. Particularly, in one of the  trifurcation junction configurations, the minimum $q_{e}$ value is observed to be $0.165$, as shown in Table~\ref{tab:angle_quality}. This is primarily due to increased localized curvature and extreme distortion near the junction introduced when the angles between the branches become very small.

\begin{table}[pos=htbp]
\centering
\caption{Scaled Jacobian statistics for different bifurcation and trifurcation junction configurations. The minimum, mean, and $10^{th}$ percentile ($P_{10}$) $q_{e}$ values are reported for each configuration.}
\label{tab:angle_quality}
\setlength{\tabcolsep}{4pt}
\begin{tabular}{lccc@{\hspace{0.7cm}}lccc}
\toprule
\multicolumn{4}{c}{\textbf{Bifurcation Junction}} &
\multicolumn{4}{c}{\textbf{Trifurcation Junction}} \\
\cmidrule(r){1-4}\cmidrule(l){5-8}
Configuration & Min($q_{e}$) & Mean($q_{e}$) & $P_{10}(q_{e})$ &
Configuration & Min($q_{e}$) & Mean($q_{e}$) & $P_{10}(q_{e})$ \\
\midrule
$(135 \degree,90 \degree,135 \degree)$ & 0.740 & 0.988 & 0.979 & $(90 \degree,90 \degree,90 \degree,90 \degree)$ & 0.657 & 0.982 & 0.954 \\
$(150 \degree,60 \degree,150 \degree)$ & 0.634 & 0.985 & 0.979 & $(135 \degree,45 \degree,45 \degree,135 \degree)$ & 0.460 & 0.979 & 0.963 \\
$(165 \degree,30 \degree,165 \degree)$ & 0.305 & 0.971 & 0.937 & $(120 \degree,60 \degree,60 \degree,120 \degree)$ & 0.512 & 0.980 & 0.962 \\
$(90 \degree,90 \degree,180 \degree)$ & 0.598 & 0.985 & 0.974 & $(60 \degree,60 \degree,60 \degree,180 \degree)$ & 0.359 & 0.975 & 0.954 \\
$(60 \degree,240 \degree,60 \degree)$ & 0.328 & 0.969 & 0.948 & $(60 \degree,210 \degree,45 \degree,45 \degree)$ & 0.357 & 0.962 & 0.905 \\
$(45 \degree,45 \degree,270 \degree)$ & 0.238 & 0.964 & 0.925 & $(150 \degree,30 \degree,30 \degree,150 \degree)$ & 0.165 & 0.974 & 0.962 \\
\bottomrule
\end{tabular}
\end{table}

\section{Results and Discussion}
We evaluate the proposed procedural modeling framework with respect to volumetric mesh generation from point cloud data, dynamic mesh updating, and suitability for FEA. We test the framework on three different plant species with increasing geometric and  topological complexity. We first demonstrate volumetric mesh generation from skeleton point clouds of mung bean (Section~\ref{subsubsec:mungbean}) and tomato (Section~\ref{subsubsec:tomato}) plants. In Section~\ref{subsec:walnut}, we demonstrate the performance of the framework for unstructured point clouds of walnut tree branches. Here, in addition to generating volumetric meshes from plant skeletons, we also estimate the branch radii from unstructured point clouds, enabling mesh generation for complex branching architectures with spatially varying radii. We then demonstrate the main advantage of the procedural framework over static mesh generation approaches: automated modeling and dynamic mesh updating during plant growth without reconstructing the entire mesh from scratch at each growth stage (Section~\ref{subsec:dynamic-growing}). Finally, in Section~\ref{subsec:diffusion-simulation}, we solve a steady-state diffusion problem on the generated meshes to demonstrate their direct suitability for FEA.

\subsection{Volumetric mesh generation from plant skeleton datasets} \label{subsec:skeleton-data-results}
We demonstrate the procedural modeling framework by generating volumetric meshes from skeleton point clouds of mung bean and tomato plants. Since these input point clouds approximate only the centerline geometry of the plants and do not provide branch-radius information, we model each branch using a circular cross-section with a prescribed constant radius along its length. Specifically, we assign a constant outer branch radius $R_{b}$ to every node $\nu_l\in \mathcal{V}_S$ in the skeleton graph $\mathcal{G}_{S}$, such that $R_l=R_{b}$, and use a constant inner-to-outer branch radius ratio $\eta$.

\subsubsection{Mung bean plants (\textit{Vigna radiata})} \label{subsubsec:mungbean}
Point clouds of several mung bean varieties were obtained via terrestrial laser scanning \cite{hadadi2026floraforge}, and skeletons defining the plant medial axes were extracted from these point clouds using the method described in \cite{Xu2007}. In all mung bean specimens, the resulting skeleton graphs contain only bifurcation junctions. Because the skeleton point clouds are relatively sparse, additional nodes are inserted along the skeleton graph edges by interpolation, so that the spacing between adjacent nodes does not exceed $1.0$ mm. A uniform outer branch radius $R_{b} = 1.5$ mm is assigned to all skeleton nodes, and $\eta = 0.5$ provides an inner branch radius of $0.75$ mm. In the volumetric parameterization for each branch, cubic B-spline basis functions are used in all three parametric directions. In the circumferential direction, we use periodic B-splines with $14$ unique control points, whereas in the radial direction, we use $4$ control points. The hex mesh vertices for each branch are evaluated on a parametric grid of $S_\theta = 28$ and $S_r = 4$ coordinates, with an axial parametric spacing of $\Delta_u = 1\,\text{mm}$. We use the hierarchical construction strategy to preserve the geometric smoothness of the parent centerline B-spline curves during branch insertion, as outlined in Algorithm \ref{alg:hierarchical}. For each branch edge $b_{\alpha}$, $n^{\alpha}$ is set equal to the number of coordinates in $\mathbf{Q}^{\alpha}$, as described in Section~\ref{subsubsec:hierarchical-construction}. We use the same parameters in the procedural framework for all the mung bean morphologies in the dataset, ranging from upright stem architectures to semi-erect, laterally sprawling branching architectures.

\begin{figure}[pos=htbp]
    \centering
\includegraphics[width=0.85\linewidth]{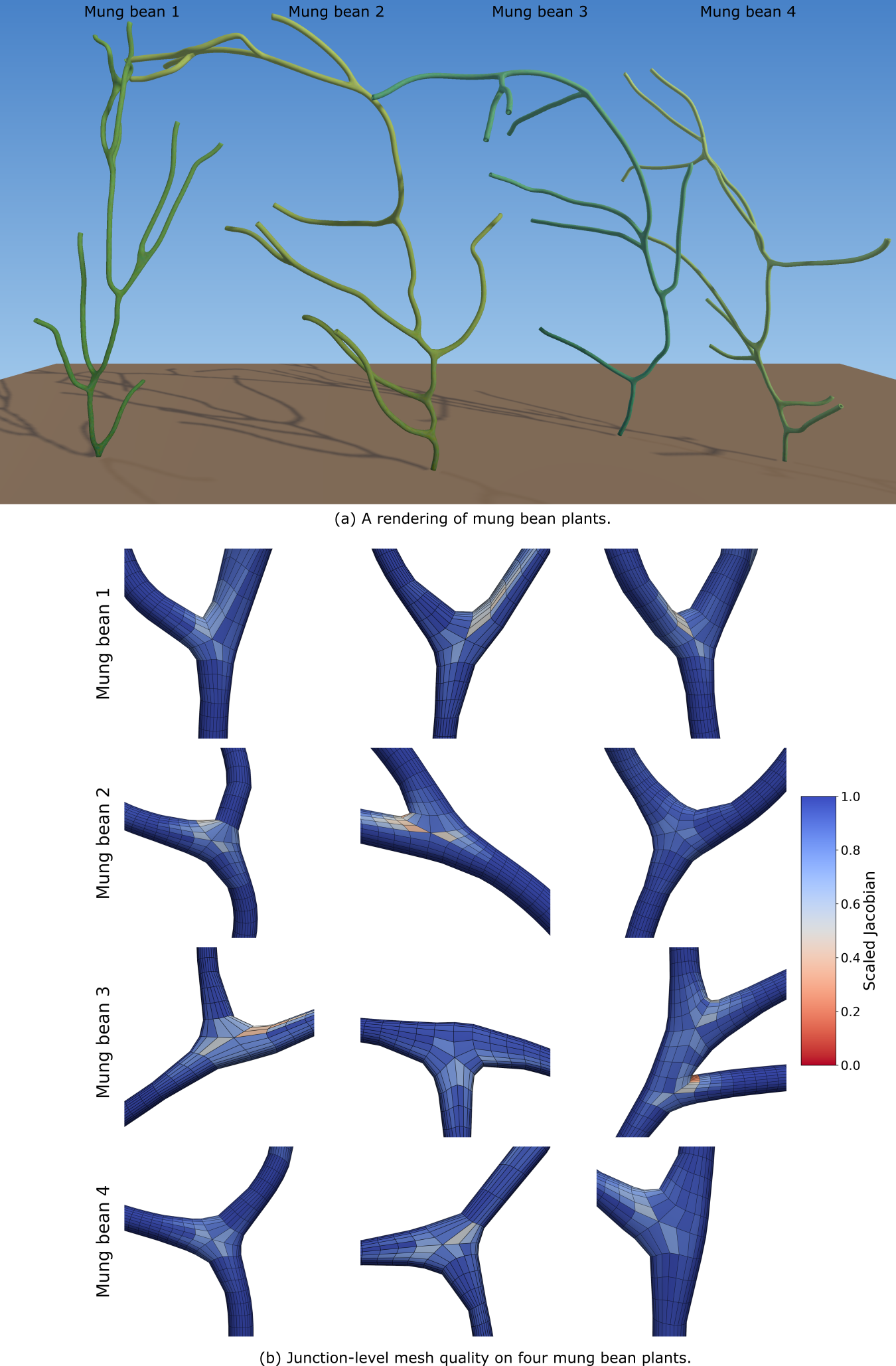}
\caption{\textbf{Mung bean plant volumetric mesh generation.} (a) Rendering of the generated volumetric hex meshes for four mung bean plant specimens (Mung bean 1--4). (b) Close-up views of three representative bifurcation junctions from each plant specimen, colored by the element-wise scaled Jacobian $q_{e}$.}
\label{fig:mungbean}
\end{figure}

Fig.~\ref{fig:mungbean}(a) shows the volumetric meshes generated from the mung bean plant skeletons, along with the corresponding close-up views of selected bifurcation junctions colored by the element-wise scaled Jacobian $q_{e}$ (Fig.~\ref{fig:mungbean}(b)). The minimum, mean, and $10^{th}$ percentile ($P_{10}$) values of $q_{e}$ are reported in Table~\ref{tab:mungbean} for each specimen. For all meshes, the mean $q_{e}$ value is greater than $0.97$, and the $P_{10}$ value is greater than $0.95$, indicating that at least $90\%$ of the mesh elements have $q_{e} > 0.95$. The minimum $q_{e}$ values are approximately between $0.1$ and $0.27$, and the lower-quality elements are localized to a small number of junctions. These junctions typically have child branches meeting the parent branch at very acute angles. The close-up views of these junctions show visually smooth parent-child branch interfaces without any kinks on the surface, consistent with the continuity conditions enforced during junction blending.

\begin{table}[pos=htbp]
\centering
\begin{minipage}{0.48\textwidth}
\centering
\begin{tabular}{lccc}
\toprule
Specimen & Min($q_{e}$) & Mean($q_{e}$) & $P_{10}$($q_{e}$) \\
\midrule
Mung bean 1 & 0.2625 & 0.9773 & 0.9542 \\
Mung bean 2 & 0.1383 & 0.9802 & 0.9622 \\
Mung bean 3 & 0.1042 & 0.9816 & 0.9746 \\
Mung bean 4 & 0.1651 & 0.9788 & 0.9567 \\
\bottomrule
\end{tabular}
\caption{Scaled Jacobian statistics for mung bean plants.}
\label{tab:mungbean}
\end{minipage}\hfill
\begin{minipage}{0.48\textwidth}
\centering
\begin{tabular}{lccc}
\toprule
Specimen & Min($q_{e}$) & Mean($q_{e}$) & $P_{10}$($q_{e}$) \\
\midrule
Tomato 1 & 0.1598 & 0.9525 & 0.9109 \\
Tomato 2 & 0.3044 & 0.9531 & 0.9088 \\
Tomato 3 & 0.3893 & 0.9546 & 0.9141 \\
Tomato 4 & 0.2516 & 0.9509 & 0.9026 \\
\bottomrule
\end{tabular}
\caption{Scaled Jacobian statistics for tomato plants.}
\label{tab:tomato}
\end{minipage}
\end{table}

\subsubsection{Tomato plants (\textit{Solanum lycopersicum})}
\label{subsubsec:tomato}
\begin{figure}[pos=htbp]
    \centering   \includegraphics[width=0.95\linewidth]{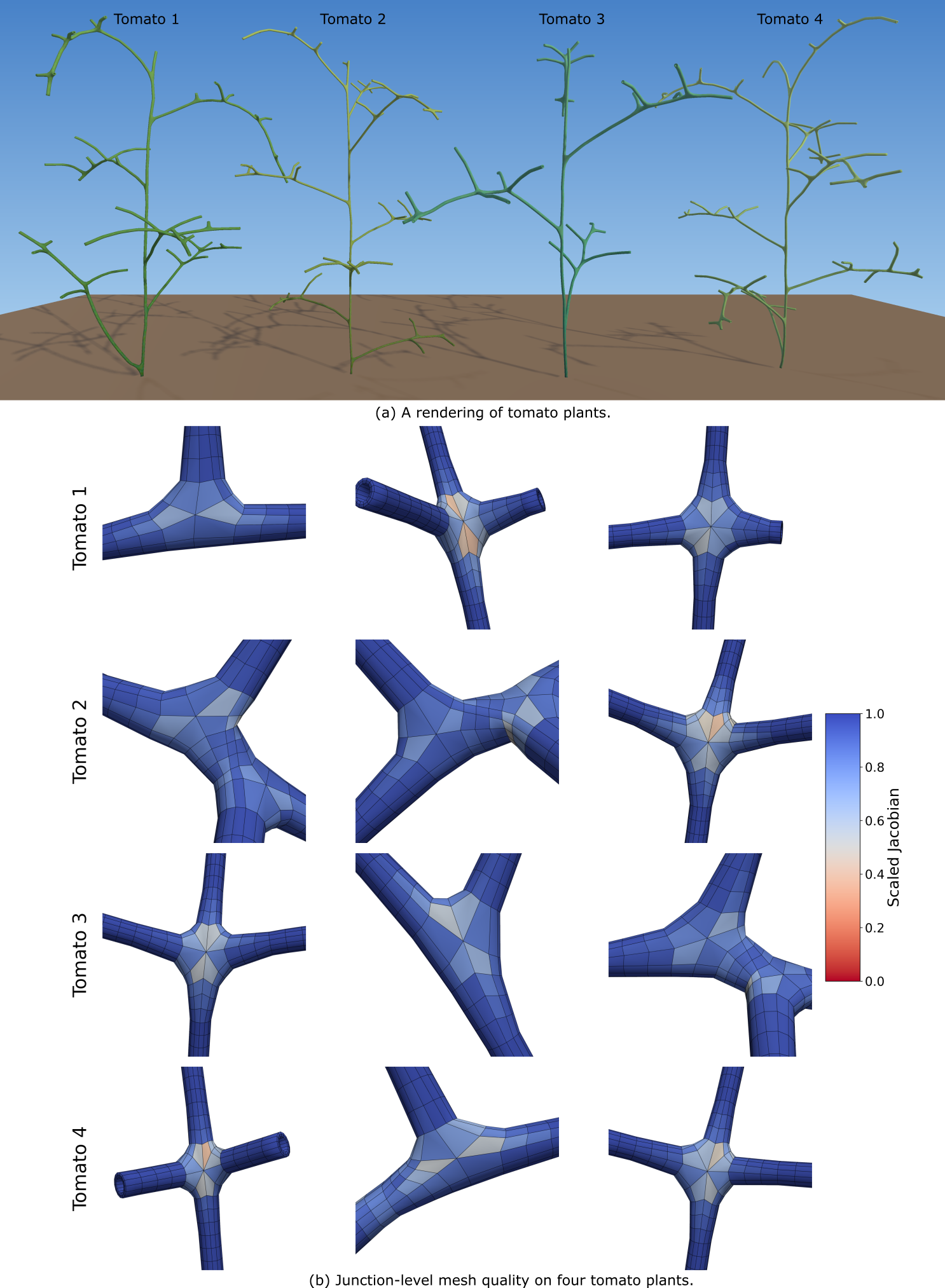}
    \caption{\textbf{Tomato plant volumetric mesh generation.} (a) Rendering of the generated volumetric hex meshes for four tomato plant specimens (Tomato 1--4). (b) Close-up views of three representative junctions from each plant model, colored by the element-wise scaled Jacobian value $q_{e}$.}
    \label{fig:tomato}
\end{figure}
We next apply the procedural framework to four skeleton point clouds of tomato plants from the TomatoWUR dataset \cite{van2025tomatowur}. The selected specimens have dataset serial numbers 101, 336, 345, and 98, corresponding respectively to Tomato 1–4 in Fig.~\ref{fig:tomato} and Table~\ref{tab:tomato}. Compared with the mung bean specimens, the tomato specimens contain more complex branching structures. In addition to plant stems, the skeletons also include large leaf-vein structures and smaller vein branches that attach near the distal ends of larger leaf branches. The knot insertion operation used in the hierarchical construction strategy (Section~\ref{subsubsec:hierarchical-construction}) requires nondegenerate parametric intervals on both sides of the child-branch attachment point on the parent B-spline curve. Specifically, if a child branch attaches to the parent B-spline centerline curve $\mathbf{C}^{\alpha}(u)$ of branch $b_{\alpha}$ at parameter $\bar{u}_{\delta}$, the construction requires two adjacent parameters, $\bar{u}_{\delta}^{-}$ and $\bar{u}_{\delta}^{+}$, such that $\bar{u}_{\delta}^{-}<\bar{u}_{\delta}<\bar{u}_{\delta}^{+}$. When the attachment point lies too close to the distal end of the parent B-spline curve, $u=1$, there is insufficient parametric space to define a distinct neighboring parameter $\bar{u}_{\delta}^{+}$ satisfying $\bar{u}_{\delta}<\bar{u}_{\delta}^{+}<1$. Therefore, the hierarchical construction strategy is not suitable for this dataset because several smaller vein branches attach near the distal ends of their parent branches. We thus use the global construction strategy described in Section~\ref{subsubsec:global-construction} to generate the B-spline centerline curves for the complete branch graph. 

We increase the node density in each skeleton graph to achieve a maximum spacing of $1.0$ mm between adjacent nodes. A constant outer branch radius $R_{b} = 2$ mm is assigned to all the skeleton nodes, and $\eta$ is set to a constant value of $0.5$, resulting in an inner branch radius of $1$ mm. The volumetric parameterization for each branch uses cubic B-spline basis functions in all three parametric directions. The number of unique control points in the circumferential direction is set to $14$, and the number of control points in the radial direction is set to $4$. The number of axial control points is determined from the branch length using the global construction strategy (Algorithm~\ref{alg:global}). The hex mesh vertices are evaluated with $S_\theta = 14$ and $S_r = 3$ parametric coordinates in each cross-section, and with an axial parametric spacing of $\Delta_u = 2\,\text{mm}$. Fig.~\ref{fig:tomato} shows the volumetric meshes for the tomato specimens, together with close-up views of selected junctions colored by the element-wise scaled Jacobian $q_{e}$. As shown in Table~\ref{tab:tomato}, the mean $q_{e}$ value for each mesh is approximately $0.95$, and $P_{10}(q_{e})$ is approximately $0.91$, both of which are slightly lower than the corresponding values obtained for the mung bean meshes. The reduction in mesh quality relative to the mung bean dataset is consistent with the higher branching density and greater variation in branch lengths of the tomato specimens. Moreover, a larger number of junctions increases the number of mesh elements affected by junction blending, which can locally reduce mesh quality. However, since each junction only affects a small number of elements, the overall mesh quality decreases very slightly.

\subsection{Walnut tree mesh generation with spatially varying branch radii}
\label{subsec:walnut}
The meshes generated for the mung bean and tomato plants in  Section~\ref{subsec:skeleton-data-results} use a constant outer branch radius for all branches within each specimen, independent of branch order. Real plants, particularly trees, exhibit substantial variation in branch radius across branch orders and along individual branches. Thus, the proposed framework must also support the blending of incident branches with different radii at junctions. We demonstrate this capability using three walnut tree specimens from the tree dataset \cite{dobbs2023smart}, which provides unstructured 3D point clouds of branching structures together with their corresponding ground-truth skeletons. At each node $\nu_l$ on the skeleton graph, we estimate the outer radius of cross-section $R_{l}$ from the neighboring points in the unstructured point cloud. To compute $R_l$, we first identify the points lying on the outer branch surface. A $k$-nearest-neighbor graph of the point cloud is constructed, and points with one-sided neighborhoods are classified as outer-surface points. The radius $R_l$ at each node is then computed as the average distance from $\nu_{l}$ to its $k$ nearest outer-surface points. Finally, the radii $R_{l}$ are interpolated along the fitted centerline B-spline curve $\mathbf{C}^{\alpha}(u)$ of each branch to obtain the outer branch radius $h_{k}$ at each Greville abscissa $\xi_{k}$ when constructing the control lattice in Eq.~\eqref{eq:cp_construction}. In this way, the spatially varying branch radius along each centerline B-spline curve is incorporated directly into the volumetric parameterization, without requiring control-point optimization. The sparse plant skeletons provided in the dataset are resampled to a maximum internodal spacing of $1\,\text{mm}$, yielding a denser skeleton point cloud for more accurate B-spline centerline curve fitting. The centerline curves for the complete branch graph are constructed using the hierarchical construction strategy described in Section~\ref{subsubsec:hierarchical-construction}, preserving the geometric smoothness of the parent centerline B-spline curve each time a child branch is attached to it. The volumetric parameterization uses cubic basis functions in all three parametric directions, with $14$ unique control points in the circumferential direction and $4$ control points in the radial direction. The hex mesh vertices are computed by evaluating each B-spline volume on a cross-sectional parametric grid of size $S_\theta = 42$ and $S_r = 4$, with the axial parametric spacing $\Delta_u$ set per specimen based on its branch-length distribution. 

\begin{figure}[pos=htbp]
    \centering
\includegraphics[width=0.87\linewidth]{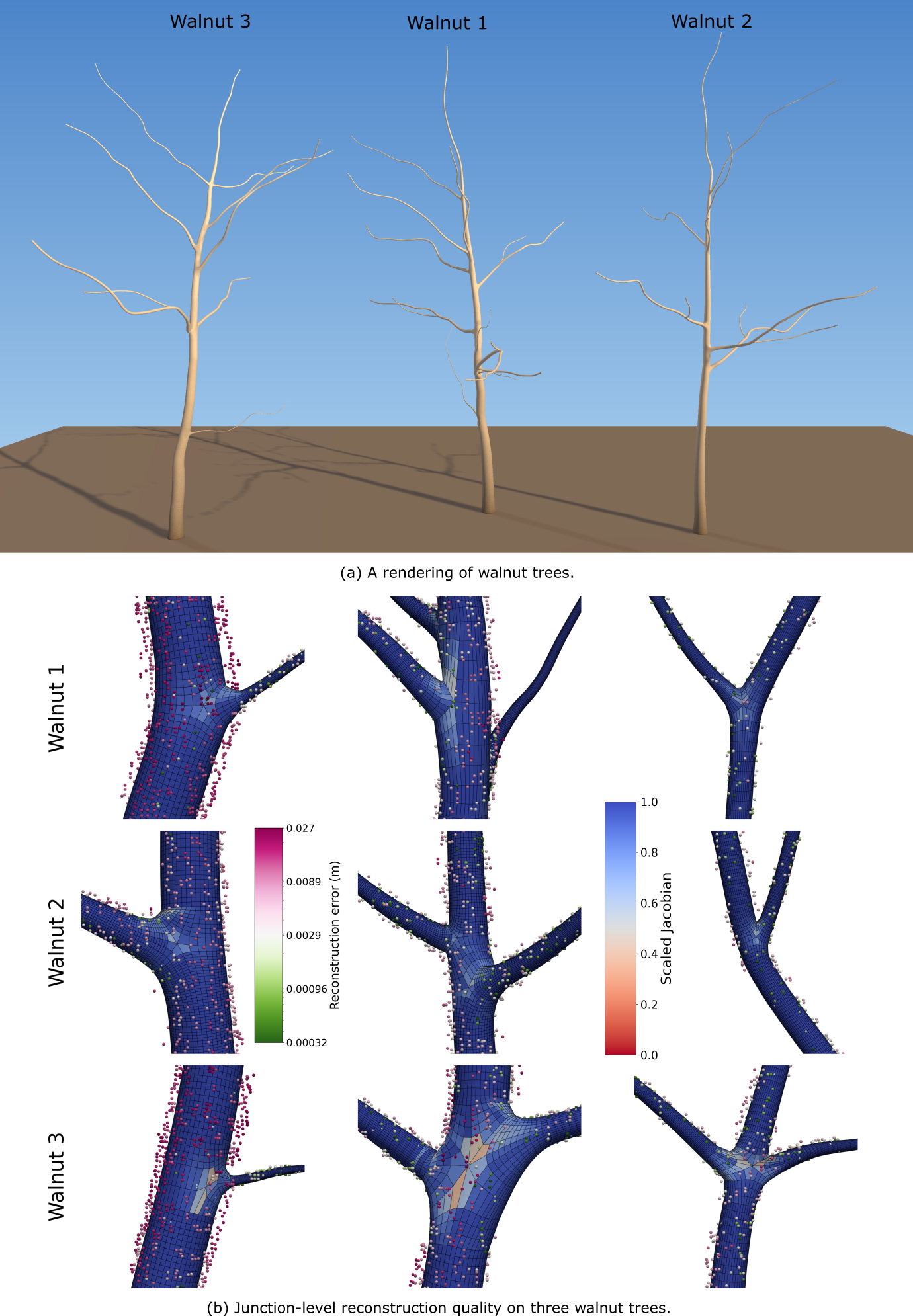}
    \caption{\textbf{Walnut tree volumetric mesh generation.} (a) Volumetric hex mesh reconstructions of three walnut trees (Walnut 1--3). (b) Three representative junction regions from each tree, colored by the element-wise scaled Jacobian. The input point cloud is overlaid on the reconstructed volumetric mesh and colored by the shortest Euclidean distance to the mesh vertices.}
    \label{fig:walnut}
\end{figure}

The reconstructed volumetric meshes of the walnut tree specimens are shown in Fig.~\ref{fig:walnut}(a), with close-up views of representative junction regions in Fig.~\ref{fig:walnut}(b). The input unstructured point cloud is overlaid on each mesh to visualize reconstruction accuracy, with each point colored by its reconstruction error, defined as the shortest Euclidean distance from the point to the mesh. Table~\ref{tab:walnut_reconstruction_error} summarizes the reconstruction error statistics across all tree specimens. We report a mean error value of less than $3$ mm and a maximum error value not exceeding $38.6$ mm across all the specimens. As reported in Table \ref{tab:walnut}, the meshes for walnut tree specimens achieve the highest overall mesh quality among the three datasets, with mean $q_{e}$ and $P_{10}(q_{e})$ both greater than $0.99$. This improvement relative to the meshes obtained in Section~\ref{subsec:skeleton-data-results} is due to the finer circumferential discretization ($S_\theta = 42$ for walnut, versus $28$ for mung bean and $14$ for tomato specimens), together with a larger number of mesh elements per branch generated in the taller walnut trees. Because the mesh elements with lower quality remain confined to a localized region at each junction, increasing the number of elements per branch subsequently increases the proportion of high-quality elements away from the junctions and thus improves the overall mesh quality. The mesh elements corresponding to the minimum $q_{e}$ values remain localized to only a small number of junction regions where branches meet at acute angles.

\begin{table}[pos=htbp]
\centering
\begin{minipage}{0.48\textwidth}
\centering
\begin{tabular}{lccc}
\toprule
Specimen & Min($q_{e}$) & Mean($q_{e}$) & $P_{10}(q_{e})$ \\
\midrule
Walnut 1 & 0.1181 & 0.9917 & 0.9941 \\
Walnut 2 & 0.3922 & 0.9943 & 0.9953 \\
Walnut 3 & 0.1052 & 0.9926 & 0.9953 \\
\bottomrule
\end{tabular}
\caption{Scaled Jacobian statistics for walnut trees.}
\label{tab:walnut}
\end{minipage}\hfill
\begin{minipage}{0.48\textwidth}
\centering
\begin{tabular}{lccc}
\toprule
Specimen & Min Error & Mean Error & Max Error\\
\midrule
Walnut 1 & 0.1 & 2.7 & 38.6 \\
Walnut 2 & 0.0 & 2.9 & 35.7 \\
Walnut 3 & 0.2 & 2.9 & 31.8 \\
\bottomrule
\end{tabular}
\caption{Reconstruction error statistics (in mm) for walnut trees.}
\label{tab:walnut_reconstruction_error}
\end{minipage}
\end{table}

\begin{table}[pos=H]
\centering
\small
\setlength{\tabcolsep}{6pt}
\renewcommand{\arraystretch}{1.15}
\begin{tabular}{l l l @{\hskip 30pt} l l l @{\hskip 30pt} l l l}
\hline
Specimen & Elements & Nodes & Specimen & Elements & Nodes & Specimen & Elements & Nodes \\
\hline
Mung bean 1 & 57,456 & 77,136  & Tomato 1 & 37,884 & 57,588 & Walnut 1 & 354,690 & 474,288 \\
Mung bean 2 & 71,064 & 95,332  & Tomato 2 & 43,904 & 66,708 & Walnut 2 & 692,118 & 923,712 \\
Mung bean 3 & 62,076 & 83,244  & Tomato 3 & 26,320 & 39,954 & Walnut 3 & 292,824 & 391,320 \\
Mung bean 4 & 94,164 & 126,132 & Tomato 4 & 45,920 & 69,840 &          &        &         \\
\hline
\end{tabular}
\caption{Hexahedral mesh statistics in terms of number of elements and vertices for the reconstructed specimens of each dataset.}
\label{tab:mesh_sizes}
\end{table}

\subsection{Dynamic plant growth modeling}
\label{subsec:dynamic-growing}
Modeling plant growth and development is a dynamic process in which the meshes must robustly adapt to extending and bifurcating branches that continuously modify the underlying plant topology. The computational cost of static mesh generation pipelines can compound over multiple growth stages, thus becoming extremely prohibitive for the dynamic modeling of complex branching architectures. In our procedural framework, we first represent the plant centerline geometry as a graph and subsequently fit it to smooth B-spline centerline curves. A tensor-product B-spline volume is then constructed around the centerline curve  of each branch and finally assembled to generate a hex mesh after junction-blending operations. This sequence of steps allows plant growth to be modeled using local graph updates. Newly introduced branch segments are first converted to B-spline branch volumes, and junction blending is carried out in a local manner while retaining the hex mesh generated in the previous growth stage. This provides a scalable approach for dynamic volumetric modeling of growing plants, enabling local mesh updates without constructing geometry from scratch.

\begin{figure}[pos=htbp]
\centering
\includegraphics[width=0.85\linewidth]{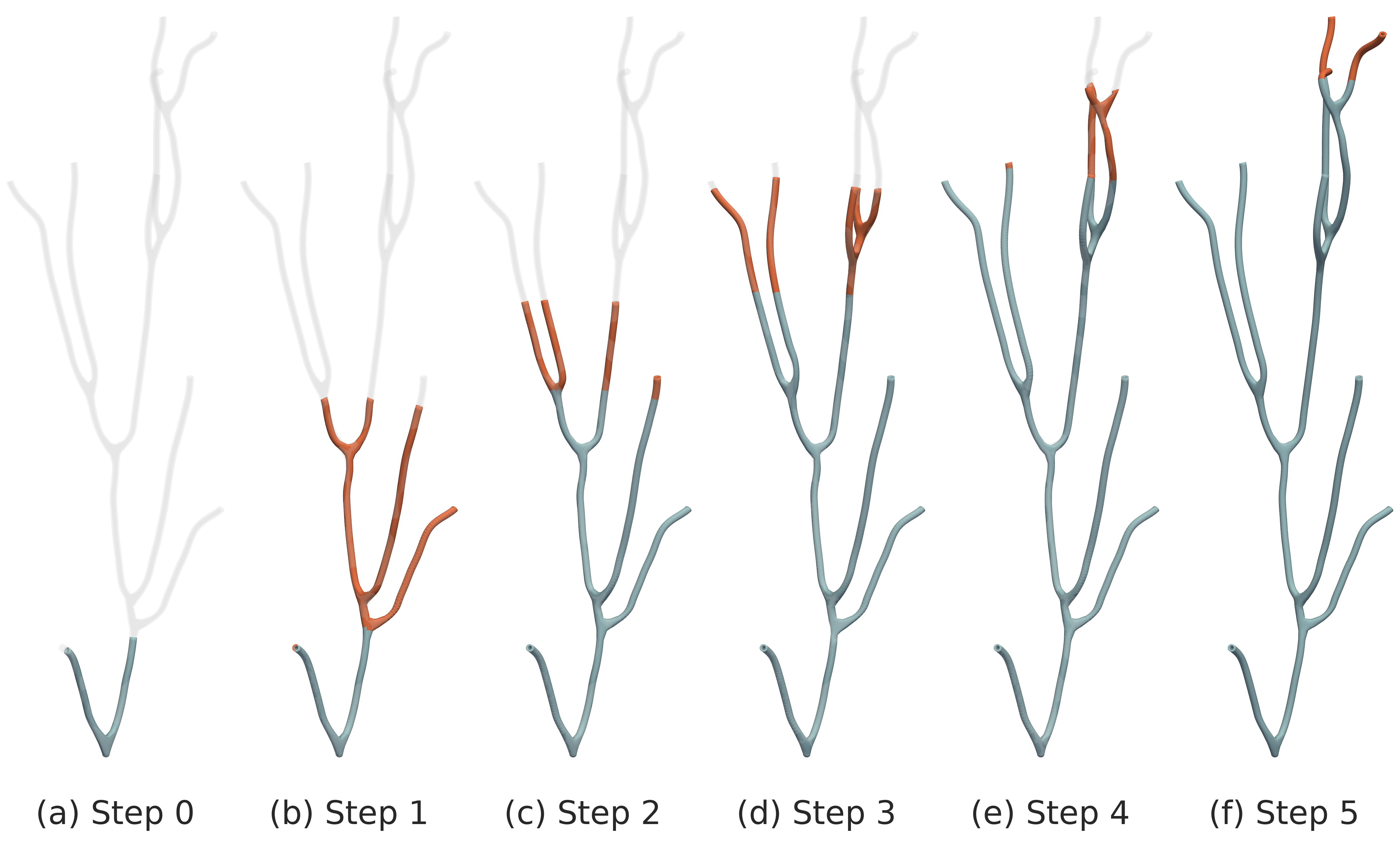}
    \caption{\textbf{Volumetric reconstruction of a mung bean plant specimen at six representative growth stages via dynamic modeling.} The meshes are generated by BFS-based frontier expansion of the skeleton graph from the root node. Newly added branches at each stage are shown in orange, while the plant geometry reconstructed in the previous stages is shown in blue. The fully mature plant geometry is overlaid as a gray silhouette.}
    \label{fig:growth_study}
\end{figure}

We demonstrate dynamic mesh updating using the Mung bean 3 specimen (Section~\ref{subsubsec:mungbean}) by simulating growth through a breadth-first search (BFS) traversal of the skeleton graph, starting from the root node and moving systematically according to the branching order. At each growth stage, only the part of the skeleton that has grown so far, referred to as the active skeleton subgraph, is considered for volumetric mesh generation. The active skeleton subgraph contains all nodes whose BFS distance from the root node is less than or equal to the current growth level. Fig.~\ref{fig:growth_study} shows the generated volumetric hex meshes at six representative growth stages. The hex mesh generated at  previous growth stages is shown in blue, whereas newly added branches at the current growth stage are shown in orange. As reported in Table~\ref{tab:growth_study}, the hex meshes generated at all the growth stages have positive scaled Jacobian values, indicating that there are no inverted mesh elements. The mean $q_{e}$ value remains at least $0.9729$ across all growth stages, indicating that the local mesh updates at each stage preserve global mesh quality. Table~\ref{tab:growth_study_times} compares the computation time required to rebuild the full plant geometry at each growth stage with the time required for dynamic updating by locally adding the new branches.  We demonstrate that reconstructing the entire plant geometry from scratch requires $0.096$ seconds at the final growth stage (Step 5). In contrast, the dynamic approach only requires between $0.03$ and $0.05$ seconds to generate the volumetric mesh at each growth stage after the initial step (Step 0). This confirms that the framework decouples local growth updates from global branching complexity, providing a scalable framework for dynamic plant modeling.

\begin{table}[pos=htbp]
\centering
\begin{minipage}{0.48\textwidth}
\centering
\captionsetup{width=0.95\linewidth}
\begin{tabular}{lccc}
\toprule
Growth Stage & Min($q_{e}$) & Mean($q_{e}$) & $P_{10}(q_{e})$ \\
\midrule
Step 0 & 0.5307 & 0.9748 & 0.9402 \\
Step 1 & 0.3879 & 0.9767 & 0.9517 \\
Step 2 & 0.2511 & 0.9756 & 0.9547 \\
Step 3 & 0.2412 & 0.9729 & 0.9450 \\
Step 4 & 0.2945 & 0.9751 & 0.9524 \\
Step 5 & 0.3427 & 0.9760 & 0.9559 \\
\bottomrule
\end{tabular}
\caption{Scaled Jacobian statistics for plant growth modeling.}
\label{tab:growth_study}
\end{minipage}\hfill
\begin{minipage}{0.48\textwidth}
\centering
\captionsetup{width=0.95\linewidth}
\begin{tabular}{lcc}
\toprule
Growth Stage & Full Plant (s) & Dynamic Update (s) \\
\midrule
Step 0 & 0.007 & 0.007 \\
Step 1 & 0.059 & 0.052 \\
Step 2 & 0.063 & 0.034 \\
Step 3 & 0.089 & 0.043 \\
Step 4 & 0.091 & 0.029 \\
Step 5 & 0.096 & 0.030 \\
\bottomrule
\end{tabular}
\caption{Computation time (in seconds) for full plant reconstruction \\ and dynamic mesh updating at each growth stage.}
\label{tab:growth_study_times}
\end{minipage}
\end{table}


\subsection{Diffusion simulation in volumetric meshes}
\label{subsec:diffusion-simulation}
To evaluate the analysis-suitability of  the generated volumetric meshes, we solved a steady-state diffusion equation on each plant geometry. The purpose of these simulations is to demonstrate the direct solver-readiness of the volumetric meshes generated by the proposed procedural modeling framework. We thus verify that the generated hex meshes are conforming and have valid element orientations. This is important for downstream simulations of plant biomechanics and transport phenomena with volumetric meshes, where poor mesh quality could affect the reliability of FEA. 
\begin{figure}[pos=htbp]
    \centering
\includegraphics[width=0.85\linewidth]{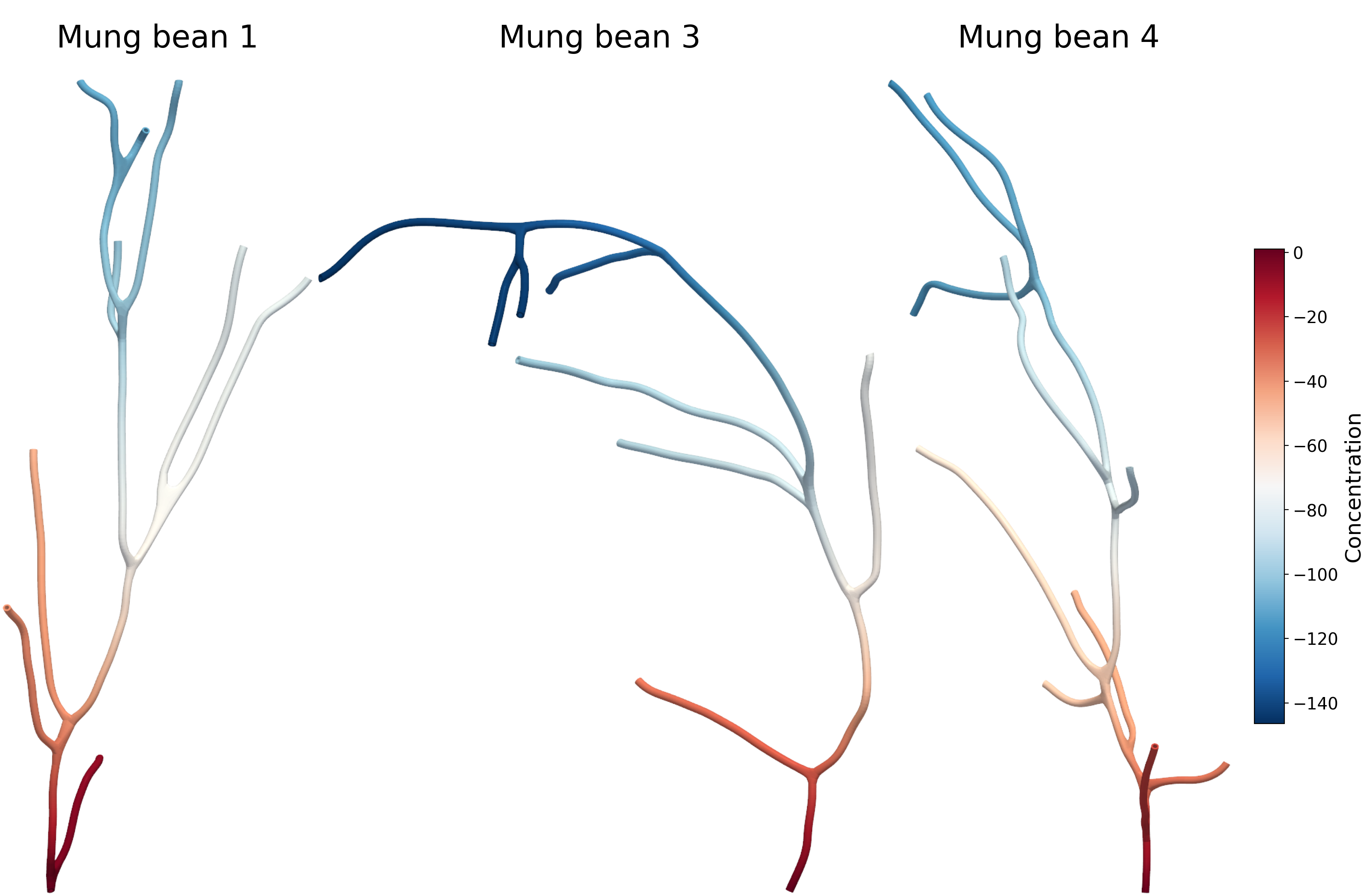}
    \caption{\textbf{Steady-state diffusion simulation on mung bean volumetric meshes.} The computed concentration field is shown for a diffusion problem with a source prescribed at the root boundary cross-section $\Gamma_{\text{root}}$ and terminal conditions applied at the leaf cross-sections $\Gamma_{\text{leaf}}$ for three mung bean plant specimens. The smooth and monotonic variation of the solution across multiple branching orders and blended junctions demonstrates the direct analysis suitability and solver-readiness of the generated hex meshes.}
    \label{fig:mungbean-sim}
\end{figure}

We define a steady-state nutrient concentration field $c(\mathbf{x})$ over the 3D plant domain $\Omega \in \mathbb{R}^3$. We denote the unique branch incident to
the root node $\nu_{\mathrm{root}}$, which is the proximal node
 in the branch graph $\mathcal{G}_{B}$, as $b_{\zeta} \in \mathcal{E}_B$. The B-spline volume $\mathbf{V}^{\zeta}$ is defined for the branch $b_{\zeta}$. A concentration $c(\mathbf{x})=1$ is prescribed on the root (source) cross-section $\Gamma_{\mathrm{root}}$, defined as the proximal ($u=0$) face of $\mathbf{V}^{\zeta}$:
\[
  \Gamma_{\mathrm{root}}
  = \bigl\{\, \mathbf{V}^{\zeta}(r,\theta,0)
    \;:\; r\in[0,1],\ \theta\in[0,1] \,\bigr\}.
\] This cross-section is referred to as the source boundary cross-section. A constant outward flux is prescribed on the terminal leaf (sink) cross-sections, defined as the distal ($u=1$) faces of all leaf branches. Let $\mathcal{I} = \bigl\{\, \alpha \;:\; b_\alpha \in \mathcal{E}_B,\
    \nu_\alpha \in \mathcal{V}_B^{\mathrm{leaf}} \,\bigr\}$ denote the index set of leaf branches. We then define the terminal leaf cross-sections as \[
  \Gamma_{\mathrm{leaf}}
  = \bigcup_{\alpha \in \mathcal{I}}
    \bigl\{\, \mathbf{V}^{\alpha}(r,\theta,1)
    \;:\; r\in[0,1],\ \theta\in[0,1] \,\bigr\}.
\] The lateral branch walls are the inner and outer cylindrical surfaces of every
branch volume, which are the faces at $r=0$ (inner) and $r=1$ (outer) of each
$\mathbf{V}^{\alpha}$:
\[
  \Gamma_{\mathrm{wall}}
  = \bigcup_{b_\alpha \in \mathcal{E}_B}
    \Bigl(
      \bigl\{\, \mathbf{V}^{\alpha}(0,\theta,u) \;:\; \theta\in[0,1],\ u\in[0,1] \,\bigr\}
      \;\cup\;
      \bigl\{\, \mathbf{V}^{\alpha}(1,\theta,u) \;:\; \theta\in[0,1],\ u\in[0,1] \,\bigr\}
    \Bigr).
\] The governing problem is modeled as the Laplace equation, with a Dirichlet boundary condition at $\Gamma_{\mathrm{root}}$ and an inhomogeneous Neumann (flux) boundary condition at $\Gamma_{\mathrm{leaf}}$. A homogeneous Neumann (no flux) boundary condition is applied at $\Gamma_{\mathrm{wall}}$. The diffusion problem is thus given as
\begin{equation}
  \begin{aligned}
    \nabla^2 c(\mathbf{x}) &= 0    && \text{in } \Omega, \\
    c(\mathbf{x})          &= 1    && \text{on } \Gamma_{\mathrm{root}}, \\
    \nabla c(\mathbf{x})\cdot\mathbf{n} &= -100 && \text{on } \Gamma_{\mathrm{leaf}}, \\
    \nabla c(\mathbf{x})\cdot\mathbf{n} &= 0    && \text{on } \Gamma_{\mathrm{wall}},
  \end{aligned}
\end{equation}
where $\mathbf{n}$ is the outward unit normal defined on the domain boundary $\partial \Omega= \Gamma_{\mathrm{root}} \,\cup\, \Gamma_{\mathrm{leaf}} \,\cup\, \Gamma_{\mathrm{wall}}$. Although the steady-state diffusion problem does not completely capture the complexity of the nutrient transport process, it provides a useful test for ensuring the smooth propagation of the concentration solution field through multiple branching levels and across the blended junction regions. 
\begin{figure}[pos=htbp]
    \centering\includegraphics[width=0.9\linewidth]{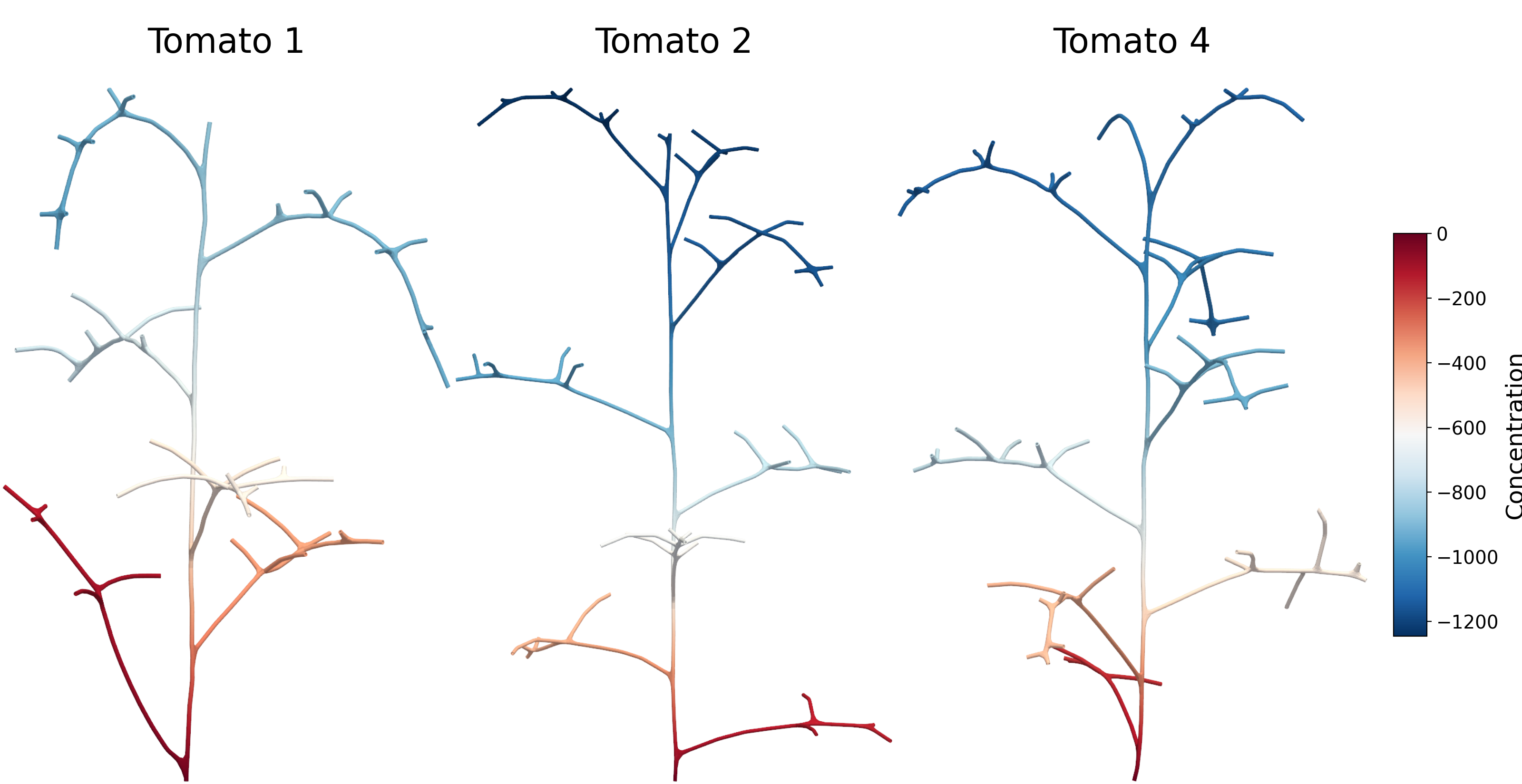}\caption{\textbf{Steady-state diffusion simulation on tomato volumetric meshes.} The computed concentration field is shown for a diffusion problem with a source prescribed at the root boundary cross-section $\Gamma_{\text{root}}$ and terminal conditions applied at the leaf cross-sections $\Gamma_{\text{leaf}}$ for three tomato plant specimens. The smooth and monotonic variation of the solution across multiple branching orders and blended junctions demonstrates the direct analysis suitability and solver-readiness of the generated hex meshes.}
\label{fig:tomato-sim}
\end{figure}

\begin{figure}[pos=htbp]
    \centering
\includegraphics[width=0.9\linewidth]{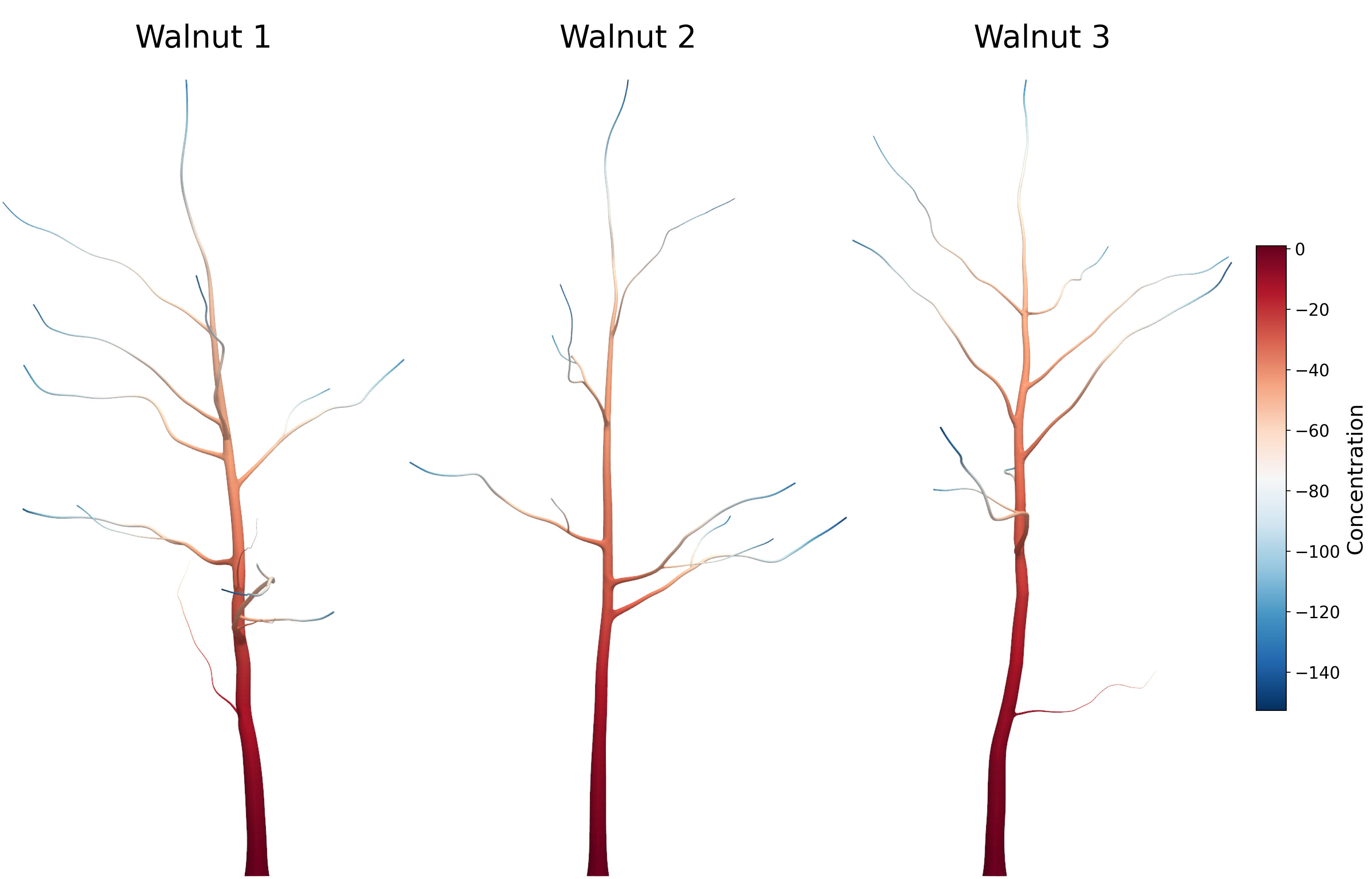}\caption{\textbf{Steady-state diffusion simulation on walnut tree volumetric meshes.} The computed concentration field is shown for a diffusion problem with a source prescribed at the root boundary cross-section $\Gamma_{\text{root}}$ and terminal conditions applied at the leaf cross-sections $\Gamma_{\text{leaf}}$ for three tree-scale architectures featuring spatially varying branch radii. The smooth and monotonic variation of the solution across multiple branching orders and blended junctions demonstrates the direct analysis suitability and solver-readiness of the generated hex meshes.}
\label{fig:walnut-sim}
\end{figure}
We perform diffusion simulations on the hex meshes generated from three mung bean, tomato, and walnut tree specimens (Figs.~\ref{fig:mungbean-sim}--~\ref{fig:walnut-sim}). The generated meshes span different levels of branching complexity, including meshes with uniform and spatially varying radii. The solver converges for every mesh, indicating that the meshes are valid for physics-based simulations.  Across all specimens, the concentration field $c(\mathbf{x})$ decreases monotonically from the root cross section toward the terminal leaf cross-sections and remains continuous across all branch junction regions. These results demonstrate the capability of the proposed procedural volumetric modeling framework to not only generate valid meshes but also to support their direct use in FEA. The solver-readiness of the volumetric modeling framework highlights its advantage over surface reconstruction methods, which often require additional mesh generation and cleanup pipelines to be carried out before physics-based simulations can be performed.

\section{Conclusion and Future Work}
We present a novel procedural framework that converts plant skeleton point clouds into  analysis-suitable hexahedral meshes for plant branching structures. The pipeline extracts a branch graph from the skeleton, fits a B-spline centerline curve to each branch, constructs a cylindrical tensor-product B-spline volume along each centerline curve, and enforces smooth junction continuity across incident branch volumes at each junction via control point blending operations to generate the hex meshes. Mesh generation at several angle configurations for bifurcation and trifurcation junctions shows that mesh quality deteriorates only at extremely acute angles while remaining free of element inversions and intersections in all cases. In highly complex branching architectures, localized element inversions can occur near the junctions. However, these can be resolved by increasing the number of control points or by increasing the pruning radius near junctions to allow more room for junction blending. We demonstrate the generalizability of the framework across different plant species by generating analysis-suitable plant geometries with consistently good mesh quality. We also demonstrate that plant growth can be modeled dynamically by extending the skeleton graph while adding new branches at each growth stage and subsequently carrying out blending at the newly added junctions. The resulting meshes are solver-ready, as evidenced by a steady-state diffusion simulation in which the concentration field remains smooth across every junction for all the generated plant geometries. Together, these results demonstrate that the framework bridges the gap between surface-based plant reconstruction and analysis-ready procedural volumetric mesh generation pipelines. Several directions remain for future work. Our framework currently focuses on bifurcation and trifurcation junctions; however, some plant architectures exhibit higher-valence junctions. Extending the junction blending operations to such junctions would enable the modeling of more complex branching architectures. Moreover, while the present work uses B-spline volumes as a scaffold for generating conforming hexahedral meshes, we plan to develop a complete B-spline-based volumetric parameterization of the entire plant geometry for isogeometric analysis.

\section*{Acknowledgements}
 A. Moola and A. Pawar would like to acknowledge funding in part from the National Science Foundation under award number DMS-2533961 and the Translational AI Center at Iowa State University.
\bibliographystyle{elsarticle-num}

\bibliography{my-bib}

\end{document}